\documentclass[preprintnumbers,twocolumn,article,amsmath,amssymb,floatfix,9pt,prd,superscriptaddress,nofootinbib]{revtex4-1}

\usepackage[utf8,latin1]{inputenc}
\usepackage{graphicx}
\usepackage{dcolumn}
\usepackage[dvipsnames]{xcolor}
\usepackage[T1]{fontenc}

\usepackage{mathrsfs}  
\usepackage{cases}
\usepackage{bm}
\usepackage{academicons}
\usepackage{mathtools, nccmath}
\usepackage{fancyhdr}
\usepackage{tikz,xcolor}

\usepackage{tensor}
\usepackage[normalem]{ulem}
\usepackage{lipsum}
\usepackage{soul}
\usepackage{cancel}
\usepackage{stackengine,scalerel}
\usepackage{mathabx}
\usepackage{hyperref}
\usepackage{tabularx}
\usepackage{makecell}
\hypersetup{colorlinks, linkcolor={red},citecolor={blue},urlcolor={blue}}  

\AtBeginDocument{\fontsize{8}{11}\selectfont}

\renewcommand{\arraystretch}{1.2}

\definecolor{lime}{HTML}{A6CE39}
\DeclareRobustCommand{\orcidicon}{
	\begin{tikzpicture}
	\draw[lime, fill=lime] (0,0) 
	circle [radius=0.16] 
	node[white] {{\fontfamily{qag}\selectfont \tiny ID}};
	\draw[white, fill=white] (-0.0625,0.095) 
	circle [radius=0.007];
	\end{tikzpicture}
	\hspace{-2mm}
}

\fancypagestyle{plain}{%
  \fancyhf{}
  \fancyfoot[C]{\iffloatpage{}{\thepage}}
  }
\foreach \x in {A, ..., Z}{%
	\expandafter\xdef\csname orcid\x\endcsname{\noexpand\href{https://orcid.org/\csname orcidauthor\x\endcsname}{\noexpand\orcidicon}}
}

\begin{document}

\title[Dymnikova-Schwinger quantum-corrected slowly rotating wormholes: Photon and spinning particle dynamics]{Dymnikova-Schwinger quantum-corrected slowly rotating wormholes: Photon and spinning particle dynamics}

\author{A. Errehymy\orcidA{}}
\email{abdelghani.errehymy@gmail.com}
\affiliation{Astrophysics Research Centre, School of Mathematics, Statistics and Computer Science, University of KwaZulu-Natal, Private Bag X54001, Durban 4000, South Africa}
\affiliation{Center for Theoretical Physics, Khazar University, 41 Mehseti Str., Baku, AZ1096, Azerbaijan}

\author{Y. Khedif\orcidB{}}
\email{youssef.khedif@gmail.com}
\affiliation{Department of Physics, Faculty of Sciences A\''{i}n Chock, Laboratory of Mechanics and High Energy Physics, University Hassan II, P.O. Box 5366, 20100, Maarif Casablanca, Morocco}

\author{M. Daoud\orcidC{}}
\email{m$_{}$daoud@hotmail.com}
\affiliation{Department of Physics, Faculty of Sciences, Ibn Tofail University, P.O. Box 133, Kenitra 14000, Morocco}
\affiliation{Abdus Salam International Centre for Theoretical Physics, Miramare, Trieste 34151, Italy}

\author{B. Turimov\orcidD{}}
\email[]{bturimov@astrin.uz}
\affiliation{Ulugh Beg Astronomical Institute, Astronomy St. 33, Tashkent 100052, Uzbekistan}
\affiliation{Engineering school, Central Asian University, Milliy bog Str.264, Tashkent, 111221,Uzbekistan}

\author{M. A. Khan\orcidE{}}
\email[]{mskhan@imamu.edu.sa}
\affiliation{Department of Mathematics and Statistics, College of Science, Imam Mohammad Ibn Saud Islamic University (IMSIU), Riyadh 11566, Saudi Arabia}

\author{S. Usanov\orcidF{}}
\email[]{sm.usanov@kiut.uz}
\affiliation{Kimyo International University in Tashkent, Shota Rustaveli Str. 156, Tashkent 100121, Uzbekistan}

\date{\today} 

\begin{abstract}
{\footnotesize This work studies light propagation near slowly rotating traversable wormholes supported by a quantum-inspired matter source. The model is based on the Dymnikova density profile, viewed as a gravitational analogue of the Schwinger mechanism, which yields a smooth, non-singular core. Quantum effects are included through the generalized uncertainty principle (GUP), introducing a minimal length scale while preserving regularity. Within a stationary and axisymmetric framework, we construct rotating wormhole solutions sustained by the GUP-corrected Dymnikova-Schwinger profile. The geometry satisfies key conditions such as asymptotic flatness and the flare-out requirement, and incorporates rotational features like frame dragging. We then examine photon motion via null geodesics. Both rotation and quantum corrections modify the photon sphere structure, with rotation producing a splitting between co-rotating and counter-rotating trajectories. This results in small asymmetries in photon paths and the shadow. These results provide a novel and consistent framework to probe quantum-gravity imprints in strong-field optics.
}
\end{abstract}

\maketitle
\textbf{Introduction:}
Wormholes, often called ``\textit{Einstein-Rosen bridges},'' are hypothetical tunnels in spacetime that could connect two far-apart regions. The idea was first hinted at by Flamm \cite{flamm1916beitrage} in 1916, who showed that the spatial part of the Schwarzschild solution could describe a kind of tunnel between distant points. Later, Einstein and Rosen \cite{Einstein:1935tc} proposed a model in which physical space is represented by two identical sheets joined by a bridge, effectively treating a particle as the connection between them. Conceptually, a wormhole acts as a shortcut: it could link locations separated by billions of light-years through a passage that might span just a few meters. The term ``wormhole'' itself was coined by Misner and Wheeler \cite{Misner:1957mt} in 1957, who also explored the possibility of tiny, charge-carrying wormholes. The first models of traversable wormholes appeared in 1973, when Ellis \cite{Ellis:1973yv} and Bronnikov \cite{Bronnikov:1973fh} independently showed that such structures could, in principle, allow passage without encountering singularities. Morris and Thorne \cite{Morris:1988cz, Morris:1988tu} later revisited the Ellis wormhole, using it as a tool to illustrate general relativity. They pointed out that keeping a wormhole open requires exotic matter---material that violates the weak energy condition \cite{Morris:1988cz, Morris:1988tu}. Building on this idea, Visser \cite{Visser:1989kh} proposed ways to design traversable wormholes while minimizing the need for exotic matter, suggesting that a traveler could pass through a wormhole without ever entering regions containing it. More recently, Cl\'{e}ment \cite{Clement:1980yx, Clement:1983ib, Clement:1983ic} developed exact solutions describing stationary, axisymmetric multi-wormholes within the Einstein-Maxwell-scalar field framework. These rotating wormholes show asymptotic behaviors similar to NUT-like spacetimes, revealing new possibilities for complex wormhole geometries.

Although wormholes have not yet been observed, Zhou and his co-workers \cite{Zhou:2016koy} explored potential signatures that could reveal the presence of an astrophysical rotating Ellis wormhole---a massive, compact object. By analyzing the iron line profile in the X-ray reflected spectrum of a thin accretion disk surrounding such a wormhole, they identified distinctive features that could, in principle, differentiate it from a Kerr black hole. This suggests that detecting a wormhole may become feasible in the near future. In a related line of research, specific observables for the Teo wormhole have been examined in terms of timelike and lightlike geodesics \cite{Tsukamoto:2014swa, Tsukamoto:2015hta}. These studies, which include high-energy particle collisions and the collisional Penrose process, provide tools that could help distinguish a rotating traversable wormhole from other compact objects---should such structures actually exist. Moreover, understanding these observables is essential for a comprehensive analysis of spin precession within the Teo wormhole framework.

In this paper, we investigate the rotating Teo wormhole \cite{Teo:1998dp}, a stationary and axially symmetric solution that generalizes the well-known Morris-Thorne construction \cite{Morris:1988tu, Morris:1988cz}. While this spacetime provides an appealing theoretical extension of static wormhole models, it necessarily violates the null energy condition \cite{Tsukamoto:2014swa}, and identifying such an object observationally remains an open problem. The bending of light by wormholes was first analyzed in the context of the Ellis geometry by Chetouani and Cl\'{e}ment \cite{Chetouani:1984qdm}. Since that early study, gravitational lensing in wormhole backgrounds has attracted sustained attention. Detailed analyses of light deflection in Ellis spacetime, including both weak- and strong-field regimes, were carried out by Tsukamoto \cite{Tsukamoto:2016zdu, Tsukamoto:2012zz, Tsukamoto:2012xs}. Nakajima and Asada examined its lensing characteristics further \cite{Nakajima:2012pu}, while Bhattacharya and Potapov developed analytical approaches to compute the deflection angle \cite{Bhattacharya:2010zzb}. Subsequent investigations expanded the scope to microlensing and retro-lensing phenomena \cite{Abe:2010ap, Tsukamoto:2017edq}, as well as to strong deflection effects in Janis-Newman-Winnicour and Ellis wormholes \cite{Tsukamoto:2016qro}. Other works considered lensing within brane-world and scalar-tensor wormhole scenarios \cite{Nandi:2006ds}, wave-optics corrections \cite{Yoo:2013cia}, the possible formation of primordial wormholes in the early universe \cite{Nojiri:1999pc}, and frame-dragging effects on light propagation in rotating optical wormhole geometries \cite{Errehymy:2025psi}. Together, these studies provide a broad foundation for understanding the observational features that might one day distinguish a rotating traversable wormhole from other compact objects.

Much of the existing literature has concentrated on static wormhole geometries. However, realistic astrophysical objects are expected to possess angular momentum, making rotation an essential ingredient in any physically meaningful model. With this perspective in mind, we investigate slowly rotating traversable wormholes supported by GUP-corrected Dymnikova-Schwinger matter. The GUP provides a practical framework for incorporating quantum gravity effects into semiclassical settings \cite{Veneziano:1986zf, Gross:1987kza, Amati:1987wq, Gross:1987ar, Amati:1988tn, Kempf:1994su, Scardigli:1999jh}. By extending the standard Heisenberg uncertainty relation, it naturally leads to the emergence of a minimal length scale and, in certain formulations, an upper bound on energy or momentum. These modifications are widely regarded as generic features of a quantum theory of gravity. Importantly, the GUP has moved beyond purely formal considerations, offering predictions such as possible modifications to photon propagation in vacuum \cite{Amelino-Camelia:1997ieq} and effects that may be tested in precision laboratory experiments \cite{Das:2008kaa, Ali:2011fa}. In many treatments, the generalized commutation relations are written in full three-dimensional form, yet actual computations are frequently simplified to one spatial dimension. Only a limited number of studies go further. For example, \cite{Kempf:1994su} and later \cite{Todorinov:2018arx} formulated the GUP within a fully relativistic framework, embedding it consistently in four-dimensional spacetime (see also \cite{Battista:2024gud}).

In the early 1990s, Dymnikova proposed an interesting model of a regular black hole \cite{Dymnikova:1992ux}. The idea was inspired by an earlier suggestion by Gliner that the interior of an extremely dense object could behave like a de Sitter vacuum, preventing the formation of a curvature singularity at the center \cite{Gliner:1966cgu}. In such a scenario, the central region is replaced by a finite-energy-density core, allowing the spacetime to remain well behaved everywhere. Later investigations \cite{Dymnikova:1996gob, Ansoldi:2008jw} suggested that the density profile appearing in Dymnikova's four-dimensional solution might be interpreted as a gravitational counterpart of the Schwinger mechanism. Around the same time, other studies examined how the Schwinger effect could be modified when quantum-gravity corrections associated with the GUP are taken into account \cite{Haouat:2013yba, Ong:2020tvo}. These connections motivated further developments. In particular, some authors have recently proposed a GUP-inspired extension of Dymnikova's model and analyzed how such corrections influence black holes and wormhole geometries \cite{Errehymy:2025djk, Errehymy:2025llh, Estrada:2023pny, Alencar:2023wyf}. Building on these ideas, this work explores how light behaves in the vicinity of slowly rotating wormholes supported by a quantum-inspired matter distribution. We employ the Dymnikova density profile and interpret it, following \cite{Dymnikova:1996gob, Ansoldi:2008jw}, as the gravitational analogue of the Schwinger mechanism responsible for particle-antiparticle pair creation in vacuum. To account for possible quantum-gravity corrections, the model is further extended by introducing a fundamental minimal length through the GUP, as discussed in \cite{Haouat:2013yba, Ong:2020tvo}. This framework allows the matter source to remain regular while incorporating modifications expected at extremely small scales. Within this setup, we construct a new class of slowly rotating traversable wormhole solutions using the stationary and axisymmetric formalism introduced by Teo \cite{Teo:1998dp}. The rotating spacetime is sourced by the GUP-corrected Dymnikova-Schwinger density profile, which smooths the central region and avoids curvature singularities. From this distribution we obtain the corresponding shape function and examine the main geometric properties of the spacetime. Particular attention is given to the conditions required for a physically consistent wormhole, including asymptotic flatness, the flare-out condition at the throat, and the behavior of the gravitational potentials associated with rotation and frame dragging. The analysis then focuses on the optical properties of the resulting geometry. We study how the quantum-corrected matter profile and rotation influence the location of photon spheres and show that rotation introduces a slight separation between co-rotating and counter-rotating photon trajectories. These features directly affect observable phenomena, such as photon paths and the structure of the shadow.

\textbf{Stationary, axisymmetric space-times:}\label{Sec:II}
\textbf{Killing symmetries and coordinate structure:}
Instead of specifying the exotic matter a priori, Teo in Ref.~\cite{Teo:1998dp} starts by proposing an appropriate spacetime geometry and then applies the Einstein field equations to find the stress-energy tensor that can support it. This geometry-first approach follows the philosophy introduced by Morris and Thorne in Ref.~\cite{Morris:1988cz}. Following this reasoning, the construction begins with a stationary, axisymmetric spacetime ansatz, which forms the foundation for modeling a rotating wormhole in Ref.~\cite{Teo:1998dp}. A spacetime is called stationary if it admits a time-like Killing vector field, 
\begin{equation}
    \xi^\alpha \equiv (\partial/\partial t)^\alpha, 
\end{equation}    
    which generates invariance under time translations. Similarly, it is axisymmetric if there exists a space-like Killing vector field, 
\begin{equation}
    \psi^\alpha \equiv (\partial/\partial \varphi)^\alpha,
\end{equation}    
    associated with rotational symmetry around the $\varphi$ direction. A spacetime is both stationary and axisymmetric when it possesses the Killing vectors $\xi^\alpha$ and $\psi^\alpha$ that commute, satisfying~\cite{Wald:1984rg}
\begin{equation}
[\xi, \psi] = 0.
\end{equation}
This commutation ensures that one can choose coordinates aligned with the directions of symmetry, typically setting $x^0 = t$, $x^1 = \varphi$, and $x^2, x^3$ for the remaining coordinates~\cite{Wald:1984rg}. Consequently, the metric components are independent of $t$ and $\varphi$, giving the general form
\begin{equation}\label{s2ae2}
ds^2 = g_{\mu\nu}(x^2, x^3) dx^\mu dx^\nu.
\end{equation}
From a physical standpoint, stationary and axisymmetric spacetimes play a central role in modeling the gravitational fields of rotating black holes and stars, capturing the essential features of their exterior geometry, as discussed in Refs.~\cite{Hartle:1967he, Hartle:1968si, Thorne:1971R} and references therein.

\textbf{Canonical metric form and symmetry constraints:}
In Ref.~\cite{Thorne:1971R}, Thorne examines the characteristics of static, axisymmetric spacetimes. He begins by noting that the coordinates $(t, \varphi, x^2, x^3)$ and $(t, \varphi + 2\pi, x^2, x^3)$ correspond to the same point, reflecting the fact that $\varphi$ is an angular coordinate around the rotation axis. As a result, the angular coordinate is naturally restricted to the range $[0, 2\pi)$. Because the spacetime is axisymmetric, it must remain unchanged under the simultaneous inversion of time and azimuthal angle, i.e., $t \to -t$ and $\varphi \to -\varphi$. This symmetry forces the off-diagonal metric components $g_{t2}$, $g_{t3}$, $g_{\varphi 2}$, and $g_{\varphi 3}$ to vanish, since they would change sign under such a transformation. As a result, the line element in Eq.~(\ref{s2ae2}) simplifies to~\cite{Papapetrou:1966zz, Carter:1969zz}
\begin{equation}\label{s2ea3}
ds^2 = g_{00} dt^2 + 2 g_{01} dt d\varphi + g_{11} d\varphi^2 + g_{ij} dx^i dx^j, \quad i,j = 2,3,
\end{equation}
where the presence of $g_{01}$ encodes the familiar frame-dragging effect (see Appendix~{\color{red}A}
).

\textbf{Coordinate freedom and asymptotic flatness:}
The coordinates in this form are unique up to transformations of the type~\cite{Thorne:1971R}
\begin{equation}
\overline{x}^2 = \overline{x}^2(x^2, x^3), \quad \overline{x}^3 = \overline{x}^3(x^2, x^3),
\end{equation}
which can be used to simplify the Einstein field equations or tailor the geometry to a specific problem. Under such transformations, the components $g_{00}$, $g_{01}$, and $g_{11}$ remain invariant, satisfying~\cite{Wald:1984rg,Thorne:1971R}
\begin{equation}
g_{00} = \xi_\alpha \xi^\alpha, \quad g_{01} = \psi_\alpha \xi^\alpha, \quad \text{and}\quad g_{11} = \psi_\alpha \psi^\alpha
\end{equation}
Finally, the spacetime described by Eq.~(\ref{s2ea3}) must be asymptotically flat. This requires 
\begin{equation}
  g_{00} \to 1, \quad g_{01} \to 1/r, \quad \text{and} \quad g_{11} \to r^2 \sin^2 \theta \quad \text{as} \quad r \to \infty,  
\end{equation}
 a property crucial for properly defining the total mass and angular momentum of the system\footnote{Here, $\theta$ and $r$ are standard spherical coordinates, not necessarily identical to $x^2$ and $x^3$~\cite{Thorne:1971R}}.
\textbf{Canonical metric of a rotating wormhole with Dymnikova-Schwinger matter:}
For a rotating wormhole, one can further specialize the stationary and axisymmetric metric by taking $g_{22}=g_{33}=g_{11}/\sin^2 x^2$ and $g_{23}=0$. With this choice, the line element in Eq.~(\ref{s2ea3}) assumes the form~\cite{Teo:1998dp}
\begin{equation}
\label{s2be1}
ds^2=-N^2 dt^2+e^\mu dr^2+r^2 K^2\left[d\theta^2+\sin^2\theta(d\varphi-\omega dt)^2\right],
\end{equation}
where the functions $N$, $\mu$, $K$, and $\omega$---often referred to as the gravitational potentials---depend only on $(x^2,x^3)\equiv(\theta,r)$. As shown in Ref.~\cite{Chandrasekhar:1985kt}, this metric clearly exhibits the frame-dragging effect encoded in the term involving $\omega$, as discussed in Appendix~{\color{red}A}.

To account for quantum-gravity-inspired effects in the matter distribution, we make use of the Dymnikova-Schwinger density profile. This choice naturally smooths out the central region of the spacetime, eliminating curvature singularities, while keeping the expected behavior at large distances. It offers a physically meaningful, regular matter configuration that ties seamlessly with models of quantum-corrected compact objects and wormholes. Following \cite{Dymnikova:1996gob, Ansoldi:2008jw}, one can view the Dymnikova profile as a gravitational analogue of the Schwinger effect, where the rate of electron-positron pair creation is given by \( \Gamma \sim \exp\left(-\frac{E_c}{E}\right)\). In its standard form, the Dymnikova density decreases exponentially from the center, providing a smooth transition from the core to the asymptotic region and ensuring a regular, singularity-free geometry. In QED, a sufficiently strong uniform electric field $E$ polarizes the vacuum, generating particle pairs efficiently once the field reaches the critical value
\(E_c = \frac{\pi \hbar m_e^2}{e}\),
where $m_e$ and $e$ are the electron mass and charge. In the gravitational context, the electric field is replaced by a curvature-induced gravitational tension, scaling as \(E \sim r^{-3}, \quad E_c \sim a^{-3}\). This correspondence motivates the 4D Dymnikova-Schwinger density profile given by \( \rho(r) = \rho_0 \, e^{-(r/a)^3}\), where $\rho_0,a>0$ are constants. Building on the previous discussion, the influence of a minimal length on the Schwinger effect has been explored in \cite{Haouat:2013yba, Ong:2020tvo} using the GUP. In this framework, the electron-positron pair production rate is modified to
\begin{equation} \label{eq:GUPrate}
\Gamma \sim \exp\Bigg(-\frac{A}{E} + B(\alpha)\, E\Bigg),
\end{equation}
where $A$ and $B(\alpha)$ are constants depending on the electron's mass and charge, $E$ is the electric field, and $\alpha$ encodes the GUP correction through
\begin{equation} \label{eq:GUP}
\Delta x \, \Delta p \sim \frac{\hbar}{2} \Bigg[1 + \alpha\frac{ (\Delta p)^2}{\hbar} \Bigg], \quad \alpha = \ell^2,
\end{equation}
with $\ell$ representing the minimal length scale. Following the gravitational analogy discussed earlier, where the electric field is replaced by a curvature-induced tension, this leads to a GUP-corrected Dymnikova-Schwinger density profile:
\begin{equation} \label{GUPDensity}
\rho(r) \;\simeq\; \rho_0 \, e^{-(r/a)^3} \;\underbrace{\Biggl( 1 + \frac{\alpha a}{r^3} \Biggr)}_{\text{GUP correction}},
\end{equation}
where $a$ sets the scale of the matter distribution and $\alpha$ introduces the minimal-length correction. In this approximate form, we have used the fact that the minimal length is extremely small, $\alpha/r^2 \ll 1$ (with $r \ge r_0$), allowing a first-order expansion.

In close analogy with the Morris-Thorne construction, Teo introduces a generalized function $\mu(r,\theta)$ defined by~\cite{Teo:1998dp}
\begin{equation}
\label{s2be3}
\mu(r,\theta)=-\ln\left(1-\frac{b(r,\theta)}{r}\right),
\end{equation}
where $b(r,\theta)$ plays the role of the shape function. The radial coordinate is then restricted to $r\ge b$, with the wormhole throat located at $r=r_0$. In the nonrotating limit, Eq.~(\ref{s2be1}) smoothly reduces to the Morris-Thorne metric through the identifications 
\begin{eqnarray}
    N(r,\theta)\to e^{\Phi(r)}, \, b(r,\theta)\to b(r), \, K(r,\theta)\to1, \, \text{and} \, \omega(r,\theta)\to 0.~~
\end{eqnarray}
 Teo further assumes that all gravitational potentials remain well-behaved at the throat to avoid curvature singularities.

Building on this framework, one can specify gravitational potentials compatible with a Dymnikova-Schwinger matter distribution as
\begin{eqnarray} 
N(r,\theta) &=& K(r,\theta) = 1 + \frac{4 J^2 \cos^2\theta}{r},\label{eq:DSPotentialsFull1}\\
b(r) &=& 8\pi \int_0^r \rho(r')\, r'^2 dr', \label{eq:DSPotentialsFull2}\\
\mu(r) &=& -\ln\Big(1 - \frac{b(r)}{r}\Big),\label{eq:DSPotentialsFull3}\\
\omega(r) &=& \frac{2 J}{r^3},\label{eq:DSPotentialsFull4}
\end{eqnarray}
where $b(r)$ now acts as a shape function sourced by the Dymnikova-Schwinger matter, with the throat located at $r = r_0$. In the non-rotating limit, the metric reduces smoothly to the Morris-Thorne form:
\begin{equation}
N(r,\theta) \to e^{\Phi(r)}, \quad b(r,\theta) \to b(r), \quad K(r,\theta) \to 1, \quad \omega(r,\theta) \to 0.
\end{equation}
ensuring continuity with the standard, static wormhole geometry while incorporating a physically motivated, regular matter source.

Once the general stationary and axisymmetric line element is established, Teo proceeds to analyze some of its key properties. In particular, the function $K(r,\theta)$ is required to be positive and nondecreasing with $r$. This allows the introduction of the proper radial distance $R\equiv rK(r,\theta)$, with $\partial R/\partial r>0$, measured from the origin at a given $(r,\theta)$. Consequently, the quantity $2\pi R\sin\theta$ can be interpreted as the proper circumference of a circular ring located at that point. Another useful quantity associated with the metric in Eq.~(\ref{s2be1}) is the discriminant~\cite{Teo:1998dp}
\begin{equation}
\label{s2be2}
D^2=g^2_{t\varphi}-g_{tt}g_{\varphi\varphi}=\big(N(r,\theta)K(r,\theta)\sin\theta\big)^2,
\end{equation}
which plays a central role in identifying horizons. The function $N(r,\theta)$ acts as a redshift function, and whenever $N=0$ one has $D^2=0$, signaling the presence of an event horizon. Since wormholes must be horizonless, Teo imposes regularity conditions on the gravitational potentials along the rotation axis $\theta=0,\pi$, requiring the $\theta$-derivatives of $N(r,\theta)$, $\mu(r,\theta)$, and $K(r,\theta)$ to vanish there. These conditions ensure that the geometry remains regular and free of singular behavior on the axis.

\begin{figure*}
\begin{center}
\includegraphics[width=12.9cm,height=4.9cm]{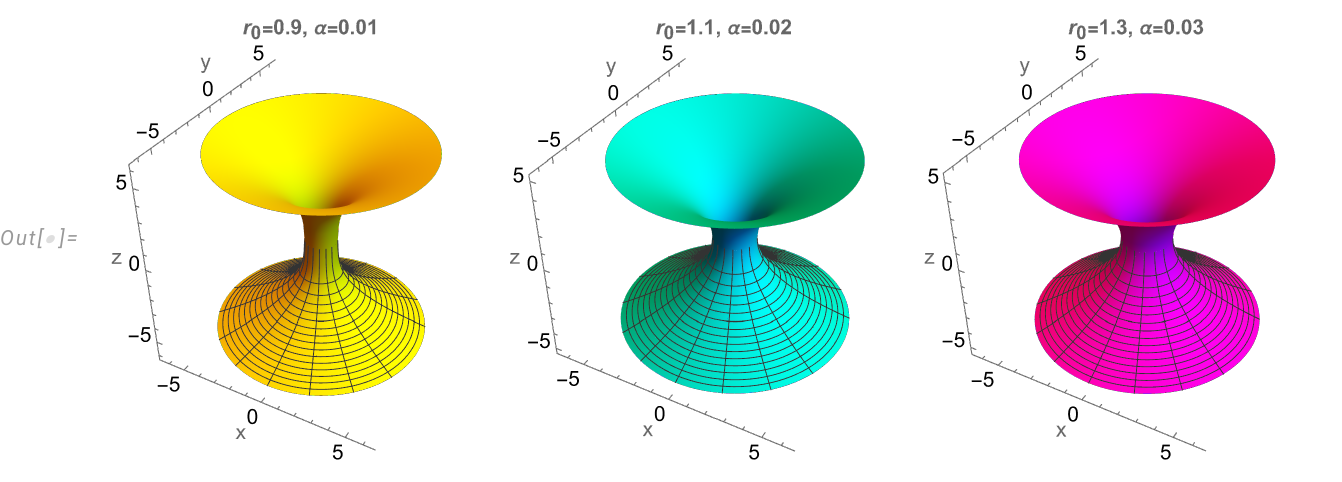}
\end{center}
\caption{{\scriptsize The three-dimensional embedding diagrams of the GUP-corrected Dymnikova-Schwinger wormhole geometry are shown for the representative parameter choices \((r_0,\alpha) = (0.9,0.01), (1.1,0.02), (1.3,0.03)\), with the scale parameter fixed at \(a=0.7\). The geometry is constructed from the shape function
\(
b(r)=\frac{r_0 f(r)}{f(r_0)}, \quad
f(r)=1-e^{-r^3/a^3}+\frac{\alpha}{a^2}\mathrm{Ei}(-r^3/a^3),
\)
where the exponential-integral term represents the GUP correction to the original Dymnikova-Schwinger profile. In each panel, the throat is positioned at \(z=0\), and both the upper and lower branches are displayed to show the complete symmetric extension of the spacetime. As the throat radius \(r_0\) increases, the minimum radius becomes wider, while larger values of the GUP parameter \(\alpha\) subtly influence the flaring-out behavior. Together, these effects illustrate how quantum-gravity corrections modify the overall structure of the wormhole.
}}\label{Fig0}
\end{figure*}

Indeed, evaluating the Ricci scalar for the metric in Eq.~(\ref{s2be1}) at the throat yields~\cite{Harko:2009xf}
\begin{eqnarray}
    \label{s2be4}
    \mathcal{R}&=&-\frac{1}{r^2 K^2}\left(\mu_{\theta\theta}+\frac{1}{2}\mu^2_\theta\right)-\frac{\mu_\theta}{Nr^2K^2}\frac{(N\sin\theta)_\theta}{\sin\theta}\nonumber\\
    &&-\frac{2}{Nr^2K^2}\frac{(N_\theta \sin\theta)_\theta}{\sin\theta}-\frac{2}{r^2K^3}\frac{(K_\theta\sin\theta)_\theta}{\sin\theta}\nonumber\\
    &&e^{-\mu}\mu_r[\ln(N r^2K^2)]_r+\frac{\sin^2\theta\omega^2_\theta}{2N^2}+\frac{2}{r^2K^4}(K^2+K^2_\theta),
    \end{eqnarray}
where subscripts denote partial derivatives with respect to $r$ and $\theta$. Potential divergences may arise from terms of the form~\cite{Teo:1998dp, Harko:2009xf}
\begin{equation}
    \label{s2be5}
    \begin{aligned}
    \mu_{\theta\theta}+\frac{1}{2}\mu^2_\theta&=\frac{b_{\theta\theta}}{r-b}+\frac{3}{2}\frac{b^2_\theta}{(r-b)^2},\quad
    \mu_\theta=\frac{b_\theta}{r-b}.
    \end{aligned}
    \end{equation}
To prevent singular behavior of the curvature scalar at the throat, one must therefore impose 
\begin{equation}
    \begin{aligned}
    b_\theta=b_{\theta\theta}=0, 
    \end{aligned}
\end{equation}
which implies that the throat is located at a constant value of $r$.

As emphasized earlier, for the metric in Eq.~(\ref{s2be1}) to genuinely represent a wormhole geometry, it must satisfy the so-called \textit{flare-out} condition at the throat. Following the procedure of Ref.~\cite{Morris:1988cz}, Teo performs an embedding of the spacetime into a higher-dimensional Euclidean space by fixing $\theta$ and considering a constant-time slice, which can be visualized as a snapshot of the entire geometry at a given instant $t$. From this embedding analysis, the flare-out condition at the throat is obtained as~\cite{Teo:1998dp}
\begin{equation}
\label{s2be6}
\frac{d^2r}{dz^2}=\frac{b-b_r r}{2b^2}>0,
\end{equation}
which coincides with the standard Morris-Thorne requirement~\cite{Morris:1988cz}. Since $b_\theta=0$, one can introduce a new proper radial coordinate defined near the throat by $l^2=r^2+b^2$, satisfying
\begin{equation}
\label{s2be7}
\frac{dl}{dr}\equiv\pm\left(1-\frac{b}{r}\right)^{-1/2}.
\end{equation}
In the immediate vicinity of the throat (to first order in $r-r_0$, with $r_0$ the throat radius), the metric in Eq.~(\ref{s2be1}) then takes the form~\cite{Teo:1998dp}
\begin{eqnarray}
    \label{s2be8}
    ds^2 = -N^2(l,\theta)\, dt^2 + dl^2 + r^2(l)\, K^2(l,\theta)\, d\Omega_\omega^2,
\end{eqnarray}
where \(d\Omega_\omega^2 \equiv d\theta^2 + \sin^2\theta\, (d\varphi - \omega(l,\theta)\, dt)^2\). Written in this way, the geometry smoothly connects two asymptotically distinct regions across the throat, whereas the coordinate $r$ becomes ill-behaved there. If the shape function is independent of $\theta$, Eq.~(\ref{s2be7}) holds globally and the coordinate $l$ spans the entire range $(-\infty,\infty)$. One may then identify the throat at $l=0$, with $l>0$ and $l<0$ corresponding to the upper and lower universes, respectively. The three-dimensional embedding diagrams of the GUP-corrected Dymnikova-Schwinger wormhole geometry are displayed in Fig.~\ref{Fig0} for representative parameter sets \((r_0,\alpha)\), while the scale parameter \(a\) is kept fixed.

Following the Morris-Thorne approach, the spacetime of Eq.~(\ref{s2be1}) can be used to compute the nonvanishing components of the stress-energy tensor by adopting a local Lorentz frame, where measurements are performed by an observer at rest with respect to the coordinates $(t,\theta,r,\varphi)$. In this frame, the relevant components are $T_{(t)(t)}$, $T_{(t)(\varphi)}$, $T_{(\varphi)(\varphi)}$, and $T_{(i)(j)}$, each having a clear physical interpretation: for instance, $T_{(t)(t)}$ corresponds to the energy density, while $T_{(t)(\varphi)}$ encodes the rotational flow of matter. Explicit expressions evaluated at the throat were obtained in Ref.~\cite{Harko:2009xf}.

To assess the energy conditions, Teo considers the null energy condition~\cite{Wald:1984rg}
\begin{equation}
\label{s2be9}
R_{\alpha\beta}\kappa^\alpha\kappa^\beta\geq 0, \quad \kappa^\alpha=\left(\frac{1}{N},-e^{-\mu/2},0,\frac{\omega}{N}\right).
\end{equation}
where $R_{\alpha\beta}$ is the Ricci tensor and $\kappa^\alpha$ is the null vector~\cite{Teo:1998dp}. Using this vector $\kappa^\alpha$, one finds~\cite{Teo:1998dp}
\begin{eqnarray}\label{s2be11}
    R_{\alpha\beta}\kappa^\alpha\kappa^\beta&=&e^{-\mu}\mu_r\frac{(rK)_r}{rK}-\frac{\omega^2_\theta\sin^2\theta}{2N^2}-\frac{1}{4}\frac{\mu^2_\theta}{(rK)^2}\nonumber\\
    &&-\frac{1}{2}\frac{(\mu_\theta\sin\theta)_\theta}{(rK)^2\sin\theta}+\frac{(N_\theta\sin\theta)_\theta}{(rK)^2N\sin\theta}<0.
\end{eqnarray}   
By suitably choosing $N$, $\mu$, and the parameters $a$, $\rho_0$, and $\alpha$, the exotic matter can be localized near the throat, allowing the null energy condition to be positive over certain angular intervals $(0,\pi)$ and ensuring a smooth, regular geometry for an infalling observer. The GUP-corrected Dymnikova-Schwinger density (\ref{GUPDensity}) provides a singularity-free matter distribution, and the corresponding shape function \(b(r)\) reads
\begin{equation} \label{b(r)2_compact}
b(r) = r_0 \, \frac{f(r)}{f(r_0)}, \quad
f(r) \equiv 1 - e^{-r^3/a^3} + \frac{\alpha}{a^2} \, \text{Ei}\Big(-\frac{r^3}{a^3}\Big)
\end{equation}
where \(f(r)\) encapsulates all exponential and GUP corrections, \(r_0\) is the throat radius, and $\text{Ei}(z)$ is the exponential integral. The ratio \(f(r)/f(r_0)\) ensures \(b(r_0) = r_0\). This construction produces a horizonless, asymptotically flat geometry with exotic matter concentrated near the throat, resulting in a quantum-inspired, physically consistent wormhole ideal for studying GUP-modified matter distributions.

\textbf{Photon motion and null geodesics in slowly rotating GUP-corrected Dymnikova-Schwinger wormholes:}\label{Sec:III}
Studying light near a rotating wormhole offers a direct glimpse into how spacetime is shaped when the source is GUP-corrected Dymnikova-Schwinger matter. In this case, the geometry is smooth and continuous, with curvature determined by the spread-out energy of the matter itself. Thanks to the GUP corrections, the energy is not sharply concentrated but smeared over a minimal length scale, resulting in a regular energy profile. This distribution directly influences the curvature, which then guides the motion of photons along null geodesics. Once rotation is introduced, even slightly, the wormhole can no longer be treated as static. Rotation couples time and angular coordinates in the metric, producing the familiar frame-dragging effect:
\begin{equation}
g_{t\phi}=-\,r^{2}\,\omega(r),
\end{equation}
where \(\omega(r)\) quantifies how local inertial frames are twisted.This twisting manifests as Lense-Thirring (\(\rm LT\)) precession, causing orbital planes to slowly rotate and producing anisotropic photon trajectories. For motion confined to the equatorial plane \((\theta = \pi/2)\) and in the slow-rotation limit, the precession frequency is directly related to the radial derivative of the rotational term:
\(\Omega_{\rm LT}(r) = -\frac{1}{2} \frac{d\omega(r)}{dr}\). In typical slowly rotating configurations, the frame-dragging velocity decreases with distance as
\(\omega(r) \simeq \frac{2J}{r^{3}}\), where \(J\) is the total angular momentum. Differentiating gives
\begin{equation}
\frac{d\omega}{dr} = -\frac{6J}{r^{4}}, \quad \text{so that} \quad \Omega_{\rm LT}(r) = \frac{3J}{r^{4}}.
\end{equation}
This shows that rotational effects are strongest near the wormhole throat and diminish rapidly with distance, leaving subtle but potentially observable signatures in lensing patterns and the wormhole shadow.

Rotation leaves its strongest imprint near the wormhole throat at \(r = r_0\). At this minimal radius, the LT precession is given by Eq. \(\Omega_{\rm LT}(r_0) = \frac{3J}{r_0^{4}}\). For slow rotations, the angular momentum is small, so the geometry remains largely governed by the shape and redshift functions, with rotation introducing only subtle kinematic effects. For example, a throat with \(r_0 = 0.25\) and \(J = 0.002\) produces a precession \(\Omega_{\rm LT} \approx 1.92\); a larger throat with \(r_0 = 0.7\) and \(J = 0.015\) reduces it to \(\Omega_{\rm LT} \approx 0.66\); and for \(r_0 = 1.1\) with \(J = 0.05\), \(\Omega_{\rm LT} \approx 0.11\). Near the throat, this rotation subtly alters photon trajectories, introducing small directional asymmetries that can influence lensing patterns and the wormhole's shadow.

\begin{figure*}
\begin{center}
\includegraphics[width=16.9cm,height=4.9cm]{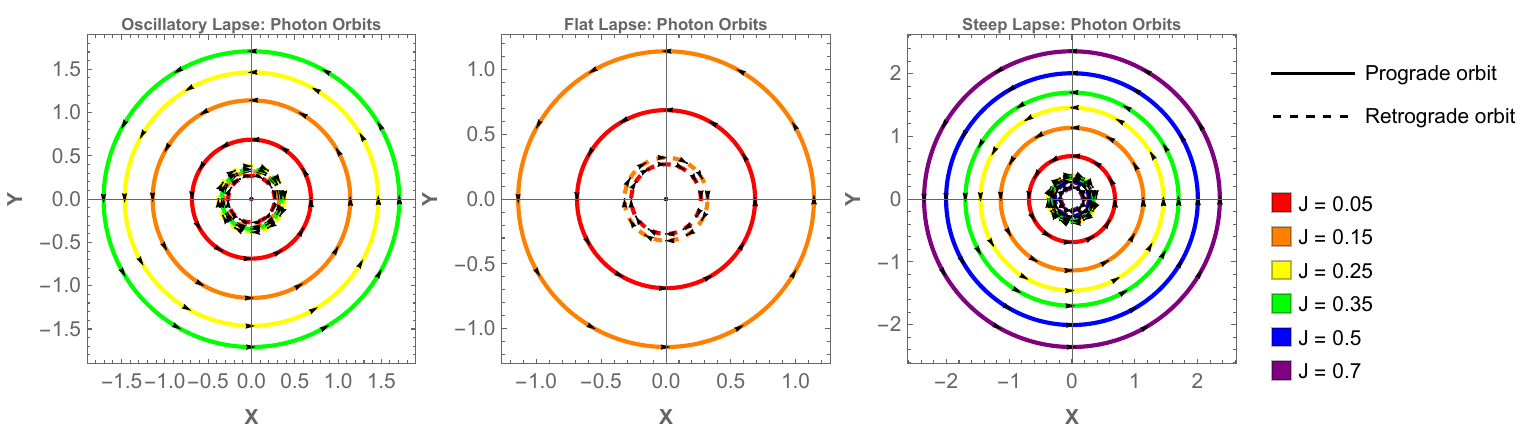}
\end{center}
\caption{{\scriptsize Photon circular orbits around a rotating wormhole are shown for three different trigonometric lapse profiles: Oscillatory, Flat, and Steep. The metric functions are controlled by the parameters \(\Phi_0 = 0.005\), \(r_0 = 0.9\), and \(\theta = 0.1\), while rotation is introduced through the frame-dragging term \(\omega(r)=2J/r^3\). The photon-sphere radius in each case is determined numerically from the corresponding extremum condition. In every panel, solid curves represent prograde motion and dashed curves represent retrograde motion for the spin values \(J = 0.05, 0.15, 0.25, 0.35, 0.5,\) and \(0.7\). The arrows mark the direction of propagation. As the spin increases, the gap between the prograde and retrograde rings steadily widens, making the frame-dragging effect clearly visible. At the same time, the overall size and distortion of the photon rings depend on the chosen lapse profile, showing how the redshift structure of the spacetime reshapes the observable photon trajectories.
}}\label{Fig1}
\end{figure*}

Now we will examine how light behaves near a wormhole. It doesn't travel in straight lines but follows the curvature of spacetime and responds to any rotation of the wormhole. For wormholes supported by GUP-corrected Dymnikova-Schwinger matter, the radial structure is dictated by the shape function \(b(r)\), while even slow rotation twists spacetime through the angular velocity \(\omega(r)\). Limiting motion to the equatorial plane (\(\theta = \pi/2\)), the metric takes the simplified form

\begin{eqnarray}
ds^2 = - N^{2} dt^2 + \frac{dr^2}{1 - b(r)/r} + r^2 (d\varphi - \omega(r) dt)^2.
\end{eqnarray}

We adopt three smooth trigonometric lapse functions to model different GUP-corrected Dymnikova-Schwinger configurations:

\begin{eqnarray}
N_{\rm Osc}(r) &=& \exp\Big[\Phi_0 \, \sin\left(\frac{r}{\tilde{r}_0}\right) \, e^{-r/\tilde{r}_c}\Big], \label{NO}\\
N_{\rm Flat}(r) &=& \exp\Big[\Phi_0 \, \big(1 - \cos\left(\frac{r}{\tilde{r}_0}\right)\big) \, e^{-r/\tilde{r}_c}\Big], \label{NF}\\
N_{\rm Steep}(r) &=& \exp\Big[\Phi_0 \, \arctan\left(\frac{r}{\tilde{r}_0}\right)\Big],\label{NS}
\end{eqnarray}
each of them finite at the throat and smoothly approaching asymptotic values at large \(r\). Here  \(\Phi_0\) sets the redshift strength, \(\tilde{r}_0\) controls how quickly it changes with radius, and \(\tilde{r}_c\) determines how fast it levels off at large distances. These slightly different profiles allow us to explore how variations in the redshift function affect photon trajectories and lensing near the wormhole, while remaining fully consistent with a smeared, GUP-corrected matter distribution.

Photon trajectories in the equatorial plane are governed by the wormhole's lapse function \(N(r)\), rotation \(\omega(r)\), and shape function \(b(r)\). Introducing conserved quantities, the energy \(\tilde{E}\) and angular momentum \(\tilde{L}\), we define the impact parameter \(\tilde{b} = \frac{\tilde{L}}{\tilde{E}}\). The radial motion follows
\begin{eqnarray}
\tilde{E} &=& N^2(r) \, \dot t + r^2 \, \omega(r) \, \dot\varphi, \\
\tilde{L} &=& r^2 \, (\dot\varphi - \omega(r) \, \dot t), \\
\dot r^2 &=& \left(1 - \frac{b(r)}{r}\right) \frac{\tilde{E}^2}{N^2(r)} \nonumber\\&\times&\Bigg[ \big(1 - \omega(r) \tilde{b} \, N(r) \big)^2 - \frac{\tilde{b}^2 N^2(r)}{r^2} \Bigg] \equiv -V(r),
\end{eqnarray}

where \(V(r)\) is the effective potential; its maxima indicate the location of photon spheres, and more generally, \(\dot r^2 + V(r) = 0\) along a trajectory.

The azimuthal evolution along the path is
\begin{eqnarray}
\frac{d\varphi}{dr} = \frac{\tilde{b} + r^2 \, \omega(r) / N^2(r)}{\sqrt{r(r - b(r)) \left[ (1 - \omega(r) \tilde{b})^2 / N^2(r) - \tilde{b}^2 / r^2 \right]}}.
\end{eqnarray}

Expanding to linear order in the rotation rate \(\omega(r)\), the total bending angle becomes
\begin{eqnarray}
\hat{\alpha} = - \pi + 2 \int_{r_\mathrm{min}}^\infty
\frac{\tilde{b} + r^2 \, \omega(r) / N^2(r)}{\sqrt{r(r - b(r)) \left[ (1 - \omega(r) \tilde{b})^2 / N^2(r) - \tilde{b}^2 / r^2 \right]}} \, dr.\qquad
\end{eqnarray}

Here, photons co-rotating with the wormhole (\(\tilde{b} > 0\)) are slightly less deflected, while counter-rotating photons (\(\tilde{b} < 0\)) experience stronger bending, reflecting the directional asymmetry induced by rotation in the wormhole geometry.

Let us now examine the special case of circular light trajectories. These orbits occur when the radial forces acting on a photon exactly balance, leading to the condition
\begin{equation}
(1 - \omega(r)\, \tilde{b})^2 = \frac{\tilde{b}^2\, N^2(r)}{r^2}.
\label{bph}
\end{equation}
When this relation is evaluated at the photon-sphere radius \(r_{\mathrm{ph}}\), it produces two possible impact parameters associated with opposite directions of motion. In the slow-rotation regime, they can be approximated as
\begin{eqnarray}
\tilde{b}_\pm(r_{\mathrm{ph}}) \simeq r_{\mathrm{ph}} N^{-1}_{\mathrm{ph}}
\pm r_{\mathrm{ph}}^{2}\,\omega_{\mathrm{ph}}\, N^{-2}_{\mathrm{ph}},
\end{eqnarray}

with \(N_{\mathrm{ph}} = N(r_{\mathrm{ph}})\) and \(\omega_{\mathrm{ph}} = \omega(r_{\mathrm{ph}})\). The two signs correspond to counter-rotating and co-rotating photons, respectively. If rotation is switched off, \(\omega(r_{\mathrm{ph}}) \to 0\), both branches collapse to the same static expression
\begin{equation}
\tilde{b}_{\mathrm{ph}} = r_{\mathrm{ph}}N^{-1}(r_{\mathrm{ph}}).
\end{equation}
The photon-sphere radius is primarily fixed by the underlying geometry, while rotation introduces a directional splitting between the two possible circular paths. In wormholes supported by GUP-corrected Dymnikova-Schwinger matter, sharper radial variations in the lapse function tend to amplify this separation, whereas smoother profiles keep the two trajectories closer together. Such a splitting, if measurable, could in principle reveal information about both the internal matter structure and the rotational state of the wormhole.

\begin{figure*}
\begin{center}
\includegraphics[width=16.9cm,height=4.9cm]{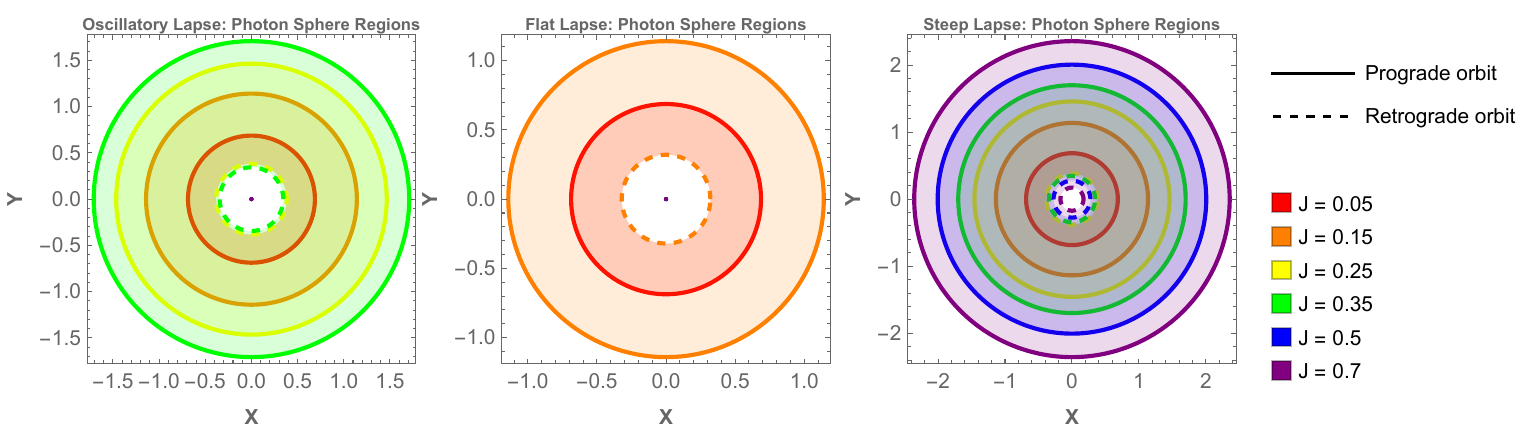}
\end{center}
\caption{{\scriptsize Photon-sphere structure for a rotating wormhole with three trigonometric lapse profiles: Oscillatory, Flat, and Steep, using \(\Phi_0 = 0.005\), \(r_0 = 0.9\), and \(\theta = 0.1\). Rotation is introduced through the frame-dragging term \(\omega(r)=2J/r^3\), and the photon-sphere radius is obtained numerically from the corresponding extremum condition for each spin value. In every panel, solid curves mark the prograde photon orbit while dashed curves indicate the retrograde one for \(J = 0.05, 0.15, 0.25, 0.35, 0.5,\) and \(0.7\). The shaded region between them highlights the allowed photon-sphere domain, making the separation between co-rotating and counter-rotating trajectories visually clear. As the spin increases, the splitting between the two radii becomes more pronounced, directly reflecting the strength of frame dragging. The overall size and thickness of the shaded regions vary from one lapse profile to another, showing how the redshift behavior of the spacetime reshapes the photon-sphere structure.
}}\label{Fig2}
\end{figure*}

Building on the three trigonometric lapse functions introduced earlier, we now examine how the shadow emerges in rotating wormholes supported by GUP-corrected Dymnikova-Schwinger matter. The boundary of the shadow is determined by the photon impact parameters evaluated at the photon-sphere radius \(r_{\mathrm{ph}}\), as given in Eq.~(\ref{bph}). This immediately shows that the shadow's overall scale and its rotational distortion are dictated jointly by the lapse function \(N(r)\) and the frame-dragging term \(\omega(r)\). Since \(N(r)\) directly enters the photon-sphere condition, each of the adopted profiles imprints a characteristic signature on light propagation. The oscillatory form \(N_{\rm Osc}(r)\), shaped by a damped sine dependence, can intensify light bending in the inner region when the redshift amplitude is appreciable, often yielding a slightly enlarged and more asymmetric shadow. The cosine-based \(N_{\rm Flat}(r)\) evolves more gradually near the throat and therefore tends to produce smoother, less distorted contours. In contrast, the arctangent profile \(N_{\rm Steep}(r)\) changes more rapidly before settling, which can accentuate the separation between prograde and retrograde photon paths while avoiding oscillatory features. If the redshift contribution becomes weak, the shadow contracts and approaches a more symmetric outline, with rotation remaining the primary source of asymmetry. Even in that limit, the observed image still depends on the viewing angle and on the emission characteristics of the surrounding region. Figs.~\ref{Fig1} and \ref{Fig2} display photon trajectories for six representative spin values \(J = 0.05, 0.15, 0.25, 0.35, 0.5\), and \(0.7\). Each panel contrasts the three lapse choices \(N_{\rm Osc}\), \(N_{\rm Flat}\), and \(N_{\rm Steep}\), illustrating how modifications of the redshift structure reshape the effective gravitational potential and shift photon orbits. Solid curves denote prograde motion, whereas dashed curves indicate retrograde trajectories. As the spin increases, prograde orbits shift outward and retrograde ones move inward, clearly reflecting the action of frame dragging.

Taken together, these results show that modest variations in the radial profile of \(N(r)\) translate into visible differences in the photon-sphere radius and orbit geometry. Consequently, the size and deformation of the shadow, as well as the structure of light rings and lensing patterns, encode information about both the wormhole's rotation and the underlying GUP-corrected matter distribution.

\begin{table*}[!htbp]
\centering
\renewcommand{\arraystretch}{0.3}
\setlength{\tabcolsep}{10pt}\scriptsize
\caption{Photon-sphere displacement and impact parameters for slowly rotating wormholes supported by GUP-corrected Dymnikova-Schwinger matter using three trigonometric lapse profiles. The shift $\delta r_\mathrm{ph}$ depends on the spin $J$ and on the radial gradient of $N(r)$, while $\tilde{b}_\pm$ encodes the rotational splitting of photon trajectories.}
\label{tab:trigo}
\resizebox{\textwidth}{!}{
\begin{tabular}{c|c|c}
\hline
{Lapse Profile} & {Model Expressions} & f{Physical Interpretation} \\
\hline

\textbf{$N_{\rm Osc}(r)$}  \\
\hline
 $N(r)$ 
& $\exp\!\Big[\Phi_0 \sin\!\left(\dfrac{r}{\tilde r_0}\right) e^{-r/\tilde r_c}\Big]$ 
& Damped oscillatory redshift near throat, \\

 $N'(r)$ 
& $N(r)\,\Phi_0 e^{-r/\tilde r_c}
\left[
\dfrac{\cos(r/\tilde r_0)}{\tilde r_0}
- \dfrac{\sin(r/\tilde r_0)}{\tilde r_c}
\right]$ 
& strong radial variation enhances sensitivity, \\

 $\delta r_\mathrm{ph}$ 
& $\dfrac{4J}{2 - r_\mathrm{ph}^{(c)} N'(r_\mathrm{ph}^{(c)})/N(r_\mathrm{ph}^{(c)})}$ 
& largest response to rotation, \\

 $\tilde{b}_\pm(r_\mathrm{ph})$ 
& $\dfrac{r_\mathrm{ph}}{N(r_\mathrm{ph})}
\Big[1 \pm \omega(r_\mathrm{ph}) r_\mathrm{ph} / N(r_\mathrm{ph}) \Big]$ 
& pronounced prograde/retrograde splitting. \\

\hline

\textbf{$N_{\rm Flat}(r)$}  \\
\hline
$N(r)$ 
& $\exp\!\Big[\Phi_0 (1 - \cos(r/\tilde r_0)) e^{-r/\tilde r_c}\Big]$ 
& Smooth, slowly varying profile, \\

$N'(r)$ 
& $N(r)\,\Phi_0 e^{-r/\tilde r_c}
\dfrac{\sin(r/\tilde r_0)}{\tilde r_0}$ 
& moderate radial gradient, \\

$\delta r_\mathrm{ph}$ 
& $\dfrac{4J}{2 - r_\mathrm{ph}^{(c)} N'(r_\mathrm{ph}^{(c)})/N(r_\mathrm{ph}^{(c)})}$ 
& mild displacement of photon sphere, \\

$\tilde{b}_\pm(r_\mathrm{ph})$ 
& $\dfrac{r_\mathrm{ph}}{N(r_\mathrm{ph})}
\Big[1 \pm \omega(r_\mathrm{ph}) r_\mathrm{ph} / N(r_\mathrm{ph}) \Big]$ 
& nearly symmetric splitting. \\

\hline

\textbf{$N_{\rm Steep}(r)$}  \\
\hline
$N(r)$ 
& $\exp\!\Big[\Phi_0 \arctan(r/\tilde r_0)\Big]$ 
& Rapid transition then saturation, \\

$N'(r)$ 
& $N(r)\,\Phi_0
\dfrac{1}{\tilde r_0 \left[1 + (r/\tilde r_0)^2\right]}$ 
& localized strong gradient, \\

$\delta r_\mathrm{ph}$ 
& $\dfrac{4J}{2 - r_\mathrm{ph}^{(c)} N'(r_\mathrm{ph}^{(c)})/N(r_\mathrm{ph}^{(c)})}$ 
& concentrated near inner region, \\

$\tilde{b}_\pm(r_\mathrm{ph})$ 
& $\dfrac{r_\mathrm{ph}}{N(r_\mathrm{ph})}
\Big[1 \pm \omega(r_\mathrm{ph}) r_\mathrm{ph} / N(r_\mathrm{ph}) \Big]$ 
& clear rotational asymmetry. \\

\hline
\end{tabular}}
\end{table*}

Rotation affects the shadow in a second, more subtle way: it slightly moves the photon-sphere radius itself. In the slow-rotation limit, the corrected position can be expressed as
\begin{eqnarray}
r_{\mathrm{ph}} \simeq r_{\mathrm{ph}}^{(c)} + \delta r_{\mathrm{ph}},
\end{eqnarray}
with the first-order correction
\begin{eqnarray}
\delta r_{\rm ph} \simeq \frac{2\, r_{\rm ph}^{(c)\,3} \, \omega(r_{\rm ph}^{(c)})}{2 - r_{\rm ph}^{(c)} \, \frac{d}{dr} \ln N(r)\big|_{r_{\rm ph}^{(c)}}}.
\end{eqnarray}
Here \(r_{\mathrm{ph}}^{(c)}\) is the photon-sphere radius in the static case, while \(\delta r_{\mathrm{ph}}\) represents the rotational correction. The denominator shows that the shift is not controlled by rotation alone; it also depends on how sharply the redshift function varies at the circular orbit. A steeper radial gradient enhances the sensitivity of the photon sphere to spin. Using the three trigonometric lapse functions introduced earlier for the GUP-corrected Dymnikova-Schwinger background, the correction takes different explicit forms:
\begin{eqnarray}
\delta r_{\mathrm{ph}}^{(\mathrm{Osc})}
&\simeq&
\frac{4J}
{2 - r_{\mathrm{ph}}^{(c)} \Phi_0 e^{-r_{\mathrm{ph}}^{(c)}/\tilde r_c}
\left[
\frac{\cos\left(r_{\mathrm{ph}}^{(c)}/\tilde r_0\right)}{\tilde r_0}- \frac{\sin\left(r_{\mathrm{ph}}^{(c)}/\tilde r_0\right)}{\tilde r_c}
  \right]},
  \\
  \delta r_{\mathrm{ph}}^{(\mathrm{Flat})}
  &\simeq&
  \frac{4J}
  {2 - r_{\mathrm{ph}}^{(c)} \Phi_0 e^{-r_{\mathrm{ph}}^{(c)}/\tilde r_c}
  \frac{\sin\left(r_{\mathrm{ph}}^{(c)}/\tilde r_0\right)}{\tilde r_0}},
  \\
  \delta r_{\mathrm{ph}}^{(\mathrm{Steep})}
  &\simeq&
  \frac{4J}
  {2 - r_{\mathrm{ph}}^{(c)} \Phi_0
  \frac{1}{\tilde r_0}
  \frac{1}{1 + \left(r_{\mathrm{ph}}^{(c)}/\tilde r_0\right)^2}}.
  \end{eqnarray}

The oscillatory profile typically shows the strongest response to rotation because of its more pronounced radial variation near the throat. The flatter cosine-based model modifies the radius more moderately, while the arctangent form produces a smoother, gradually saturating correction. In the end, the observed shadow is shaped by two linked mechanisms: the separation between co- and counter-rotating impact parameters and the displacement of the photon sphere itself. Since both effects depend on the detailed behavior of \(N(r)\), even small differences in the underlying GUP-corrected matter distribution can leave visible imprints on the size and asymmetry of the shadow.

The trigonometric lapse profiles slightly shift the photon-sphere radius away from the throat value \(r_{0}\). To keep the analysis perturbative, we write
\(r_{\mathrm{ph}} = r_{0} + \delta r,\, |\delta r| \ll r_{0}\). At leading order, the correction depends only on how rapidly the lapse function varies at the throat. Expressed directly through \(N(r)\), the displacement reads
\begin{eqnarray}
\delta r \simeq
-\frac{r_{0}^{3}}{2}
\frac{N'(r_{0})}{N(r_{0})}
\left(1-b'(r_{0})\right)^{-1}.
\end{eqnarray}

This makes the interpretation transparent: the photon sphere reacts to the local logarithmic slope \(N'/N\). A steeper radial variation produces a larger shift.

\begin{itemize}
    \item For Eq. (\ref{NO}) (Sinusoidal profile), the logarithmic derivative becomes
\begin{eqnarray}
\frac{N'_{\mathrm{Osc}}}{N_{\mathrm{Osc}}}
=
\Phi_0 e^{-r/r_c}
\left[
\frac{\cos(r/r_0)}{r_0}
-
\frac{\sin(r/r_0)}{r_c}
\right].
\end{eqnarray}
Hence,
\begin{eqnarray}
\delta r_{\mathrm{Osc}}
\simeq
-\frac{\Phi_0 r_{0}^{3}}{2\left(1-b'(r_{0})\right)}
e^{-r_{0}/r_c}
\left[
\frac{\cos(r_{0}/r_0)}{r_0}
-
\frac{\sin(r_{0}/r_0)}{r_c}
\right].
\end{eqnarray}
The oscillatory behavior is tempered by the exponential factor, so the net shift depends on the balance between periodic modulation and damping.
    \item For Eq. (\ref{NF}) (Cosine profile), we obtain

\begin{eqnarray}
\frac{N'_{\mathrm{Flat}}}{N_{\mathrm{Flat}}}
=
-\Phi_0 e^{-r/r_c}
\left[
\frac{\sin(r/r_0)}{r_0}
+
\frac{\cos(r/r_0)}{r_c}
\right].
\end{eqnarray}
This leads to
\begin{eqnarray}
\delta r_{\mathrm{Flat}}
\simeq
\frac{\Phi_0 r_{0}^{3}}{2\left(1-b'(r_{0})\right)}
e^{-r_{0}/r_c}
\left[
\frac{\sin(r_{0}/r_0)}{r_0}
+
\frac{\cos(r_{0}/r_0)}{r_c}
\right].
\end{eqnarray}
Because sine and cosine remain bounded, the displacement stays moderate and oscillatory.
    \item For Eq. (\ref{NS}) (Tangent profile), the slope simplifies to
\begin{eqnarray}
\frac{N'_{\mathrm{Steep}}}{N_{\mathrm{Steep}}}
=
\frac{\Phi_0}{r_0}
\sec^{2}\left(\frac{r}{r_0}\right).
\end{eqnarray}
Accordingly,
\begin{eqnarray}
\delta r_{\mathrm{Steep}}
\simeq
-\frac{\Phi_0 r_{0}^{3}}
{2 r_0\left(1-b'(r_{0})\right)}
\sec^{2}\left(\frac{r_{0}}{r_0}\right).
\end{eqnarray}

Since the tangent function varies more sharply, this case produces the most localized and potentially strongest response near the throat, provided the perturbative condition remains satisfied.
\end{itemize}

Overall, the structure is simple: once \(N(r)\) is specified, the photon-sphere shift follows directly from its logarithmic gradient at the throat. Gentle oscillations lead to mild corrections, while steeper profiles translate into stronger geometric adjustments.

The wormhole shadow is set by photons grazing the photon sphere and escaping to a distant observer. Using the lapse function \(N(r)\), the critical impact parameters read
\begin{eqnarray}
b_{\pm}(r_{\rm ph}) \simeq \frac{r_{\rm ph}}{N(r_{\rm ph})} \Big(1 \pm \frac{\omega(r_{\rm ph}), r_{\rm ph}}{N(r_{\rm ph})}\Big),
\end{eqnarray}
with minus for co-rotating and plus for counter-rotating photons. The shadow size is controlled by \(N(r_{\rm ph})\), while rotation introduces a small asymmetry.
\begin{itemize}
    \item Sin profile: Mild oscillations in \(N(r)\) near the throat produce small shifts in the photon sphere, giving nearly circular shadows with minimal splitting.
    \item Cos profile: Smooth, bounded variation leads to moderate photon-sphere displacement and slightly asymmetric shadows.
    \item Tan profile: Steeper radial changes enhance co-/counter-rotating separation, producing the most noticeable distortion of the shadow.
\end{itemize}

In all cases, the shadow's shape reflects the balance between the radial lapse gradient and rotational frame dragging: photons along the rotation move outward, while those against it stay closer to the throat. The sin profile yields almost symmetric rings, the cos profile gives gentle asymmetry, and the tan profile concentrates deviations, producing the sharpest distortions.

\textbf{Concluding Remarks:}\label{Sec:IV}
In this work we have explored the structure of rotating wormholes within the framework of stationary and axisymmetric space-times, focusing in particular on configurations supported by a quantum-inspired Dymnikova-Schwinger matter distribution with corrections coming from the GUP. The aim was to understand how such physically motivated matter sources can sustain a regular rotating wormhole geometry while incorporating possible minimal-length effects expected from quantum gravity.

The starting point of the analysis was the general description of stationary and axisymmetric space-times, characterized by the presence of two commuting Killing vector fields associated with time translations and rotational symmetry. These symmetries strongly constrain the form of the metric and allow it to be written in a canonical structure where all gravitational potentials depend only on the radial and polar coordinates. Within this setting, the off-diagonal component of the metric naturally appears and encodes the familiar frame-dragging phenomenon, which reflects the rotational nature of the spacetime. Requiring asymptotic flatness further ensures that the geometry behaves like ordinary flat spacetime at large distances and allows meaningful definitions of global quantities such as mass and angular momentum. Using this geometric background, we adopted the rotating wormhole metric introduced by Teo, which generalizes the Morris-Thorne construction to stationary configurations. The geometry is governed by a set of gravitational potentials that determine the redshift behavior, spatial curvature, and rotational properties of the spacetime. Regularity requirements impose important constraints on these functions, particularly near the wormhole throat and along the rotation axis. These conditions guarantee that the spacetime remains smooth and free of curvature singularities, while also preventing the appearance of event horizons. As a result, the geometry describes a genuine traversable wormhole connecting two distinct asymptotic regions.

A central element of the model is the matter source responsible for sustaining the wormhole. Instead of introducing an arbitrary exotic fluid, we considered the Dymnikova density profile, which provides a smooth and regular matter distribution. This profile can be interpreted as a gravitational analogue of the Schwinger mechanism, where particle-antiparticle pairs are produced in a strong background field. In the gravitational setting, the exponentially decaying density replaces the singular behavior typically associated with point-like sources and leads to a regular interior geometry. To account for possible quantum-gravity effects, we incorporated corrections arising from the GUP. The existence of a minimal length modifies the particle creation process associated with the Schwinger mechanism and introduces additional contributions to the density profile. These corrections are controlled by the GUP parameter and become relevant in the region close to the wormhole throat. The resulting matter distribution remains smooth while slightly altering the structure of the shape function that determines the wormhole geometry. The geometric consequences of these modifications were examined through the embedding analysis of the wormhole throat. The flare-out condition, which is essential for the existence of a wormhole, was verified and shown to hold for the constructed solutions. The embedding diagrams provide a clear visualization of the geometry, showing how the throat connects two different regions of spacetime. Variations of the model parameters demonstrate that increasing the throat radius widens the minimal surface, while the GUP parameter introduces small corrections to the flaring behavior of the geometry. The energy conditions were also considered in order to understand the role of exotic matter in this configuration. As expected, the null energy condition is violated in the immediate vicinity of the throat, which is a typical feature of wormhole solutions. However, the analysis indicates that this violation can be confined to a relatively small region around the throat, while the energy conditions may remain satisfied over larger angular intervals. This localization of exotic matter is an appealing aspect of the model, since it reduces the amount of nonclassical matter required to support the wormhole.

Taken together, the results show that the combination of the rotating wormhole geometry with a GUP-corrected Dymnikova-Schwinger matter distribution leads to a consistent and regular spacetime configuration. The geometry remains horizonless and asymptotically flat, while the matter distribution smoothly supports the throat without introducing singularities. The inclusion of minimal-length effects offers a natural way to incorporate quantum corrections into the model and slightly modifies the structure of the wormhole near its core. This framework therefore provides a useful setting for exploring how quantum-inspired matter distributions may influence the geometry of rotating compact objects. Further studies could investigate additional aspects of these spacetimes, such as particle motion, photon trajectories, gravitational lensing, or stability under perturbations. Such analyses may help clarify whether wormhole geometries of this type could produce observable signatures or provide insight into the role of quantum effects in strong gravitational fields.

In summary, the rotating wormhole model constructed here illustrates how classical geometric techniques and quantum-inspired matter sources can be combined to produce a regular and physically motivated spacetime. By incorporating both rotation and minimal-length corrections, the model contributes to a broader effort to understand how exotic geometries might emerge within realistic extensions of general relativity.

\section*{Acknowledgements:} 
This research was also funded by the Science Committee of the Ministry of Science and Higher Education of the Republic of Kazakhstan (Grant No. AP23487178).

\textbf{Conflict Of Interest statement:} 
The authors have no conflicts of interest to disclose.

\textbf{Data Availability Statement:} 
No data were created or analyzed in this study.


\appendix
	\section{The dragging of the inertial frame\label{appenA}}
	In this appendix, we follow Chandrasekhar's approach to illustrate the presence of a frame-dragging effect in the spacetime described by Eq.~(\ref{s2be1}). Accordingly, the contravariant components of the metric tensor \(g^{\mu\nu}\) can be written in matrix form\footnote{Here, we adopt the notation \(t \rightarrow 0\), \(\theta \rightarrow 2\), and \(r \rightarrow 3\).}.
	\begin{equation}
	    \label{A1}
	    (g^{\mu\nu})=\left(
        \begin{array}{cccc}
        -\frac{1}{N^2} & -\frac{\omega }{N^2} & 0 & 0 \\
        -\frac{\omega }{N^2} & -\frac{A}{K^2 N^2 r^2} & 0 & 0 \\
        0 & 0 & \frac{1}{K^2 r^2} & 0 \\
        0 & 0 & 0 & e^{-\mu } \\
        \end{array}
        \right),
	\end{equation}
	where is defined as 
	\begin{equation}
	    \label{A2}
	    A=(K r \omega -N \csc (\theta )) (K r \omega +N \csc (\theta )).
	\end{equation}
    This spacetime is associated with the following tetrad\footnote{We adopt the same notation as in Ref.~\cite{Chandrasekhar:1985kt}.}.
    \begin{equation}
        \label{A3}
        \begin{aligned}
            e_{(c)\mu}&=(-N,0,0,0),\\
            e_{(1)\mu}&=(-rK\omega\sin\theta, rK\sin\theta,0,0),\\
            e_{(2)\mu}&=(0,0,rK,0),\\
            e_{(3)\mu}&=(0,0,0,e^\frac{\mu}{2}).
        \end{aligned}
    \end{equation}
   Using the relation \(e^{\;\;\;\;\mu}_{(a)}=g^{\mu\nu}e_{(a)\nu}\), the corresponding contravariant vectors can be expressed as follows:
        \begin{eqnarray}
            e^{\;\;\;\;\mu}_{(c)}&=\left(\frac{1}{N},\frac{\omega}{N},0,0\right), \label{Aa4}\\
            e^{\;\;\;\;\mu}_{(1)}&=\left(0, \frac{1}{rK\sin\theta},0,0\right),\label{Ab4}\\
            e^{\;\;\;\;\mu}_{(2)}&=\left(0,0,\frac{1}{rK},0\right),\label{Ac4}\\
           e^{\;\;\;\;\mu}_{(3)}&=\left(0,0,0,e^{-\frac{\mu}{2}}\right).\label{Ad4}
        \end{eqnarray}
    Therefore, for the tetrad so defined, we have 
    \begin{equation}
        \label{A5}
        e^{\;\;\;\;\mu}_{(a)}e_{(b)\mu}=\eta_{(a)(b)}=\left(
        \begin{array}{cccc}
        -1 & 0 & 0 & 0\\
         0 & 1 & 0 & 0\\
         0 & 0 & 1 & 0\\
         0 & 0 & 0 & 1
        \end{array}
        \right).
    \end{equation}
    This indicates that the selected frame is Minkowskian, meaning it locally represents an \textit{inertial frame}. One can easily verify that the components of the line element (\ref{s2be1}) follow directly from the relation:
    \begin{equation}
        \label{A6}
        g_{\mu\nu}=\eta^{(a)(b)}e_{(a)\mu}e_{(b)\nu}.
    \end{equation}
    
    The components of the four-velocity take the following form
    \begin{equation}
        \label{A7}
        \begin{array}{ccc}
            u^0=\frac{dt}{d\lambda}, & u^1=\Omega u^0, & u^\alpha=v^{\alpha} u^0\ . 
        \end{array}
    \end{equation}
    Here, \(\alpha = 2, 3\), \(v^\alpha = dx^\alpha/dt\), and \(\Omega \equiv d\varphi/dt\). In the inertial frame, the components of the four-velocity are determined from the relation:
    \begin{equation}
        \label{A8}
        u^{(a)}=\eta^{(a)(b)}e_{(b)\mu}u^\mu,
    \end{equation}
    from which 
    \begin{eqnarray}
            u^{(c)}&= N u^0,\label{Aa9}\\
            u^{(1)}&= (\Omega -\omega)rK\sin\theta u^0,\label{Ab9}\\
            u^{(2)}&= rKv^2u^0,\label{Ac9}\\
            u^{(3)}&= e^\frac{\mu}{2}v^3u^0.\label{Ad9}
    \end{eqnarray}
    From the second relation in Eqs.~(\ref{Aa9})-(\ref{Ad9}), it follows that a point undergoing circular motion with angular velocity \(\Omega\) in the coordinates \((t, \varphi, \theta, r)\) will appear to rotate with angular velocity \((\Omega - \omega) r K \sin\theta, u^0\) in the inertial frame. Conversely, a point at rest in the local inertial frame (i.e., \(u^{(1)} = u^{(2)} = u^{(3)} = 0)\) will have angular velocity \(\omega\) when viewed in the coordinate frame. This nonzero \(\omega\) therefore represents the dragging of inertial frames. In an asymptotically flat spacetime, this reduces to \(\omega = 2J/r^3\).

\bibliography{references}

@book{flamm1916beitrage,
  title={Beitr{\"a}ge zur Einsteinschen gravitationstheorie},
  author={Flamm, Ludwig},
  year={1916},
  publisher={Hirzel}
}

@article{Einstein:1935tc,
    author = "Einstein, Albert and Rosen, N.",
    title = "{The Particle Problem in the General Theory of Relativity}",
    doi = "10.1103/PhysRev.48.73",
    journal = "Phys. Rev.",
    volume = "48",
    pages = "73--77",
    year = "1935"
}

@article{Misner:1957mt,
    author = "Misner, Charles W. and Wheeler, John A.",
    title = "{Classical physics as geometry: Gravitation, electromagnetism, unquantized charge, and mass as properties of curved empty space}",
    doi = "10.1016/0003-4916(57)90049-0",
    journal = "Annals Phys.",
    volume = "2",
    pages = "525--603",
    year = "1957"
}

@article{Ellis:1973yv,
    author = "Ellis, H. G.",
    title = "{Ether flow through a drainhole - a particle model in general relativity}",
    doi = "10.1063/1.1666161",
    journal = "J. Math. Phys.",
    volume = "14",
    pages = "104--118",
    year = "1973"
}

@article{Bronnikov:1973fh,
    author = "Bronnikov, K. A.",
    title = "{Scalar-tensor theory and scalar charge}",
    journal = "Acta Phys. Polon. B",
    volume = "4",
    pages = "251--266",
    year = "1973"
}

@article{Morris:1988cz, 
    author = "Morris, M. S. and Thorne, K. S.",
    title = "{Wormholes in space-time and their use for interstellar travel: A tool for teaching general relativity}",
    doi = "10.1119/1.15620",
    journal = "Am. J. Phys.",
    volume = "56",
    pages = "395--412",
    year = "1988"
}

@article{Morris:1988tu,
    author = "Morris, M. S. and Thorne, K. S. and Yurtsever, U.",
    title = "{Wormholes, Time Machines, and the Weak Energy Condition}",
    doi = "10.1103/PhysRevLett.61.1446",
    journal = "Phys. Rev. Lett.",
    volume = "61",
    pages = "1446--1449",
    year = "1988"
}

@article{Visser:1989kh,
    author = "Visser, Matt",
    title = "{Traversable wormholes: Some simple examples}",
    eprint = "0809.0907",
    archivePrefix = "arXiv",
    primaryClass = "gr-qc",
    reportNumber = "LA-UR-89-46",
    doi = "10.1103/PhysRevD.39.3182",
    journal = "Phys. Rev. D",
    volume = "39",
    pages = "3182--3184",
    year = "1989"
}

@article{Clement:1980yx,
    author = "Clement, Gerard",
    title = "{EINSTEIN-MAXWELL HIGGS SOLITONS}",
    reportNumber = "IPUC-80-1",
    doi = "10.1007/BF00758213",
    journal = "Gen. Rel. Grav.",
    volume = "13",
    pages = "747",
    year = "1981"
}

@article{Clement:1983ib,
    author = "Clement, Gerard",
    title = "{A CLASS OF STATIONARY AXISYMMETRIC SOLUTIONS OF EINSTEIN-MAXWELL SCALAR FIELD THEORIES}",
    reportNumber = "IPUC-83-6",
    doi = "10.1088/0264-9381/1/3/007",
    journal = "Class. Quant. Grav.",
    volume = "1",
    pages = "283",
    year = "1984"
}

@article{Clement:1983ic,
    author = "Clement, Gerard",
    title = "{REGULAR MULTIPARTICLE SOLUTIONS OF EINSTEIN-MAXWELL SCALAR FIELD THEORIES}",
    reportNumber = "IPUC 83-5",
    doi = "10.1088/0264-9381/1/3/006",
    journal = "Class. Quant. Grav.",
    volume = "1",
    pages = "275",
    year = "1984"
}

@article{Zhou:2016koy,
    author = "Zhou, Menglei and Cardenas-Avendano, Alejandro and Bambi, Cosimo and Kleihaus, Burkhard and Kunz, Jutta",
    title = "{Search for astrophysical rotating Ellis wormholes with X-ray reflection spectroscopy}",
    eprint = "1603.07448",
    archivePrefix = "arXiv",
    primaryClass = "gr-qc",
    doi = "10.1103/PhysRevD.94.024036",
    journal = "Phys. Rev. D",
    volume = "94",
    number = "2",
    pages = "024036",
    year = "2016"
}

@article{Tsukamoto:2014swa,
    author = "Tsukamoto, Naoki and Bambi, Cosimo",
    title = "{High energy collision of two particles in wormhole spacetimes}",
    eprint = "1411.5778",
    archivePrefix = "arXiv",
    primaryClass = "gr-qc",
    doi = "10.1103/PhysRevD.91.084013",
    journal = "Phys. Rev. D",
    volume = "91",
    number = "8",
    pages = "084013",
    year = "2015"
}

@article{Tsukamoto:2015hta,
    author = "Tsukamoto, Naoki and Bambi, Cosimo",
    title = "{Collisional Penrose process in a rotating wormhole spacetime}",
    eprint = "1503.06386",
    archivePrefix = "arXiv",
    primaryClass = "gr-qc",
    doi = "10.1103/PhysRevD.91.104040",
    journal = "Phys. Rev. D",
    volume = "91",
    pages = "104040",
    year = "2015"
}

@article{Chetouani:1984qdm,
    author = "Chetouani, Lyazid and Cl{\'e}ment, G{\'e}rard",
    title = "{Geometrical optics in the Ellis geometry}",
    doi = "10.1007/BF00762440",
    journal = "Gen. Rel. Grav.",
    volume = "16",
    number = "2",
    pages = "111--119",
    year = "1984"
}

@article{Tsukamoto:2016zdu,
    author = "Tsukamoto, Naoki and Harada, Tomohiro",
    title = "{Light curves of light rays passing through a wormhole}",
    eprint = "1607.01120",
    archivePrefix = "arXiv",
    primaryClass = "gr-qc",
    reportNumber = "RUP-16-21",
    doi = "10.1103/PhysRevD.95.024030",
    journal = "Phys. Rev. D",
    volume = "95",
    number = "2",
    pages = "024030",
    year = "2017"
}

@article{Tsukamoto:2012zz,
    author = "Tsukamoto, Naoki and Harada, Tomohiro",
    title = "{Signed magnification sums for general spherical lenses}",
    eprint = "1211.0380",
    archivePrefix = "arXiv",
    primaryClass = "gr-qc",
    reportNumber = "RUP-12-11",
    doi = "10.1103/PhysRevD.87.024024",
    journal = "Phys. Rev. D",
    volume = "87",
    number = "2",
    pages = "024024",
    year = "2013"
}

@article{Tsukamoto:2012xs,
    author = "Tsukamoto, Naoki and Harada, Tomohiro and Yajima, Kohji",
    title = "{Can we distinguish between black holes and wormholes by their Einstein ring systems?}",
    eprint = "1207.0047",
    archivePrefix = "arXiv",
    primaryClass = "gr-qc",
    reportNumber = "RUP-12-5",
    doi = "10.1103/PhysRevD.86.104062",
    journal = "Phys. Rev. D",
    volume = "86",
    pages = "104062",
    year = "2012"
}

@article{Nakajima:2012pu,
    author = "Nakajima, Koki and Asada, Hideki",
    title = "{Deflection angle of light in an Ellis wormhole geometry}",
    eprint = "1204.3710",
    archivePrefix = "arXiv",
    primaryClass = "gr-qc",
    doi = "10.1103/PhysRevD.85.107501",
    journal = "Phys. Rev. D",
    volume = "85",
    pages = "107501",
    year = "2012"
}

@article{Bhattacharya:2010zzb,
    author = "Bhattacharya, Amrita and Potapov, Alexander A.",
    title = "{Bending of light in Ellis wormhole geometry}",
    doi = "10.1142/S0217732310033748",
    journal = "Mod. Phys. Lett. A",
    volume = "25",
    pages = "2399--2409",
    year = "2010"
}

@article{Abe:2010ap,
    author = "Abe, F.",
    title = "{Gravitational Microlensing by the Ellis Wormhole}",
    eprint = "1009.6084",
    archivePrefix = "arXiv",
    primaryClass = "astro-ph.CO",
    doi = "10.1088/0004-637X/725/1/787",
    journal = "Astrophys. J.",
    volume = "725",
    pages = "787--793",
    year = "2010"
}

@article{Tsukamoto:2017edq,
    author = "Tsukamoto, Naoki",
    title = "{Retrolensing by a wormhole at deflection angles {\ensuremath{\pi}} and 3{\ensuremath{\pi}}}",
    eprint = "1701.09169",
    archivePrefix = "arXiv",
    primaryClass = "gr-qc",
    doi = "10.1103/PhysRevD.95.084021",
    journal = "Phys. Rev. D",
    volume = "95",
    number = "8",
    pages = "084021",
    year = "2017"
}

@article{Tsukamoto:2016qro,
    author = "Tsukamoto, Naoki",
    title = "{Strong deflection limit analysis and gravitational lensing of an Ellis wormhole}",
    eprint = "1607.07022",
    archivePrefix = "arXiv",
    primaryClass = "gr-qc",
    doi = "10.1103/PhysRevD.94.124001",
    journal = "Phys. Rev. D",
    volume = "94",
    number = "12",
    pages = "124001",
    year = "2016"
}

@article{Nandi:2006ds,
    author = "Nandi, Kamal Kanti and Zhang, Yuan-Zhong and Zakharov, Alexander V.",
    title = "{Gravitational lensing by wormholes}",
    eprint = "gr-qc/0602062",
    archivePrefix = "arXiv",
    doi = "10.1103/PhysRevD.74.024020",
    journal = "Phys. Rev. D",
    volume = "74",
    pages = "024020",
    year = "2006"
}

@article{Yoo:2013cia,
    author = "Yoo, Chul-Moon and Harada, Tomohiro and Tsukamoto, Naoki",
    title = "{Wave Effect in Gravitational Lensing by the Ellis Wormhole}",
    eprint = "1302.7170",
    archivePrefix = "arXiv",
    primaryClass = "gr-qc",
    reportNumber = "YITP-13-15, RUP-13-3",
    doi = "10.1103/PhysRevD.87.084045",
    journal = "Phys. Rev. D",
    volume = "87",
    pages = "084045",
    year = "2013"
}

@article{Nojiri:1999pc,
    author = "Nojiri, S. and Obregon, O. and Odintsov, S. D. and Osetrin, K. E.",
    title = "{Can primordial wormholes be induced by GUTs at the early universe?}",
    eprint = "gr-qc/9904035",
    archivePrefix = "arXiv",
    reportNumber = "NDA-FP-59",
    doi = "10.1016/S0370-2693(99)00565-1",
    journal = "Phys. Lett. B",
    volume = "458",
    pages = "19--28",
    year = "1999"
}

@article{Errehymy:2025psi,
    author = "Errehymy, Abdelghani and Guvendi, Abdullah and Gurtas Dogan, Semra and Mustafa, Omar",
    title = "{Frame-dragging and light deflection in rotating optical wormhole spacetimes}",
    doi = "10.1016/j.physletb.2025.139847",
    journal = "Phys. Lett. B",
    volume = "869",
    pages = "139847",
    year = "2025"
}

@article{Veneziano:1986zf,
    author = "Veneziano, G.",
    title = "{A Stringy Nature Needs Just Two Constants}",
    reportNumber = "CERN-TH-4397/86",
    doi = "10.1209/0295-5075/2/3/006",
    journal = "EPL",
    volume = "2",
    pages = "199",
    year = "1986"
}

@article{Gross:1987kza, 
    author = "Gross, David J. and Mende, Paul F.",
    title = "{The High-Energy Behavior of String Scattering Amplitudes}",
    reportNumber = "PUPT-1062",
    doi = "10.1016/0370-2693(87)90355-8",
    journal = "Phys. Lett. B",
    volume = "197",
    pages = "129--134",
    year = "1987"
}

@article{Amati:1987wq, 
    author = "Amati, D. and Ciafaloni, M. and Veneziano, G.",
    title = "{Superstring Collisions at Planckian Energies}",
    reportNumber = "CERN-TH-4782/87",
    doi = "10.1016/0370-2693(87)90346-7",
    journal = "Phys. Lett. B",
    volume = "197",
    pages = "81",
    year = "1987"
}

@article{Gross:1987ar, 
    author = "Gross, David J. and Mende, Paul F.",
    title = "{String Theory Beyond the Planck Scale}",
    reportNumber = "PUPT-1067",
    doi = "10.1016/0550-3213(88)90390-2",
    journal = "Nucl. Phys. B",
    volume = "303",
    pages = "407--454",
    year = "1988"
}

@article{Amati:1988tn, 
    author = "Amati, D. and Ciafaloni, M. and Veneziano, G.",
    title = "{Can Space-Time Be Probed Below the String Size?}",
    reportNumber = "CERN-TH-5207-88, SISSA-121-88-EP",
    doi = "10.1016/0370-2693(89)91366-X",
    journal = "Phys. Lett. B",
    volume = "216",
    pages = "41--47",
    year = "1989"
}

@article{Kempf:1994su,
    author = "Kempf, Achim and Mangano, Gianpiero and Mann, Robert B.",
    title = "{Hilbert space representation of the minimal length uncertainty relation}",
    eprint = "hep-th/9412167",
    archivePrefix = "arXiv",
    reportNumber = "DAMTP-94-105",
    doi = "10.1103/PhysRevD.52.1108",
    journal = "Phys. Rev. D",
    volume = "52",
    pages = "1108--1118",
    year = "1995"
}

@article{Scardigli:1999jh,
    author = "Scardigli, Fabio",
    title = "{Generalized uncertainty principle in quantum gravity from micro - black hole Gedanken experiment}",
    eprint = "hep-th/9904025",
    archivePrefix = "arXiv",
    doi = "10.1016/S0370-2693(99)00167-7",
    journal = "Phys. Lett. B",
    volume = "452",
    pages = "39--44",
    year = "1999"
}

@article{Amelino-Camelia:1997ieq,
    author = "Amelino-Camelia, G. and Ellis, John R. and Mavromatos, N. E. and Nanopoulos, Dimitri V. and Sarkar, Subir",
    title = "{Tests of quantum gravity from observations of gamma-ray bursts}",
    eprint = "astro-ph/9712103",
    archivePrefix = "arXiv",
    reportNumber = "ACT-18-97, CTP-TAMU-49-97, OUTP-97-73-P, NEIP-97-013",
    doi = "10.1038/31647",
    journal = "Nature",
    volume = "393",
    pages = "763--765",
    year = "1998"
}

@article{Das:2008kaa,
    author = "Das, Saurya and Vagenas, Elias C.",
    title = "{Universality of Quantum Gravity Corrections}",
    eprint = "0810.5333",
    archivePrefix = "arXiv",
    primaryClass = "hep-th",
    doi = "10.1103/PhysRevLett.101.221301",
    journal = "Phys. Rev. Lett.",
    volume = "101",
    pages = "221301",
    year = "2008"
}

@article{Ali:2011fa,
    author = "Ali, Ahmed Farag and Das, Saurya and Vagenas, Elias C.",
    title = "{A proposal for testing Quantum Gravity in the lab}",
    eprint = "1107.3164",
    archivePrefix = "arXiv",
    primaryClass = "hep-th",
    doi = "10.1103/PhysRevD.84.044013",
    journal = "Phys. Rev. D",
    volume = "84",
    pages = "044013",
    year = "2011"
}

@article{Todorinov:2018arx,
    author = "Todorinov, Vasil and Bosso, Pasquale and Das, Saurya",
    title = "{Relativistic Generalized Uncertainty Principle}",
    eprint = "1810.11761",
    archivePrefix = "arXiv",
    primaryClass = "gr-qc",
    doi = "10.1016/j.aop.2019.03.014",
    journal = "Annals Phys.",
    volume = "405",
    pages = "92--100",
    year = "2019"
}

@article{Battista:2024gud,
    author = "Battista, Emmanuele and Capozziello, Salvatore and Errehymy, Abdelghani",
    title = "{Generalized uncertainty principle corrections in Rastall{\textendash}Rainbow Casimir wormholes}",
    eprint = "2409.09750",
    archivePrefix = "arXiv",
    primaryClass = "gr-qc",
    doi = "10.1140/epjc/s10052-024-13656-y",
    journal = "Eur. Phys. J. C",
    volume = "84",
    number = "12",
    pages = "1314",
    year = "2024"
}

@book{Wald:1984rg,
    author = "Wald, Robert M.",
    title = "{General Relativity}",
    doi = "10.7208/chicago/9780226870373.001.0001",
    publisher = "Chicago Univ. Pr.",
    address = "Chicago, USA",
    year = "1984"
}

@article{Hartle:1967he,
    author = "Hartle, James B.",
    title = "{Slowly rotating relativistic stars. 1. Equations of structure}",
    doi = "10.1086/149400",
    journal = "Astrophys. J.",
    volume = "150",
    pages = "1005--1029",
    year = "1967"
}

@article{Hartle:1968si,
    author = "Hartle, James B. and Thorne, Kip S.",
    title = "{Slowly Rotating Relativistic Stars. II. Models for Neutron Stars and Supermassive Stars}",
    doi = "10.1086/149707",
    journal = "Astrophys. J.",
    volume = "153",
    pages = "807",
    year = "1968"
}

@article{Thorne:1971R,
  title={Relativistic stars, black holes and gravitational waves (including an in-depth review of the theory of rotating, relativistic stars).},
  author={Thorne, Kip S},
  journal={Rend. Scu. Int. Fis. Enrico Fermi 47: 237-83 (1971).},
  year={1970},
  publisher={California Inst. of Tech., Pasadena}
}

@article{Dymnikova:1996gob,
    author = "Dymnikova, I. G.",
    title = "{De Sitter-schwarzschild Black Hole: its Particlelike Core and Thermodynamical Properties}",
    doi = "10.1142/s0218271896000333",
    journal = "Int. J. Mod. Phys. D",
    volume = "05",
    number = "05",
    pages = "529--540",
    year = "1996"
}

@inproceedings{Ansoldi:2008jw,
    author = "Ansoldi, Stefano",
    title = "{Spherical black holes with regular center: A Review of existing models including a recent realization with Gaussian sources}",
    booktitle = "{Conference on Black Holes and Naked Singularities}",
    eprint = "0802.0330",
    archivePrefix = "arXiv",
    primaryClass = "gr-qc",
    reportNumber = "KUNS-2108",
    month = "2",
    year = "2008"
}

@article{Haouat:2013yba,
    author = "Haouat, S. and Nouicer, K.",
    title = "{Influence of a Minimal Length on the Creation of Scalar Particles}",
    eprint = "1310.6966",
    archivePrefix = "arXiv",
    primaryClass = "hep-th",
    doi = "10.1103/PhysRevD.89.105030",
    journal = "Phys. Rev. D",
    volume = "89",
    number = "10",
    pages = "105030",
    year = "2014"
}

@article{Papapetrou:1966zz,
    author = "Papapetrou, Achille",
    title = "{Champs gravitationnels stationnaires {\`a} sym{\'e}trie axiale}",
    journal = "Ann. Inst. H. Poincare Phys. Theor. A",
    volume = "4",
    number = "2",
    pages = "83--105",
    year = "1966"
}

@article{Carter:1969zz,
    author = "Carter, Brandon",
    title = "{Killing horizons and orthogonally transitive groups in space-time}",
    doi = "10.1063/1.1664763",
    journal = "J. Math. Phys.",
    volume = "10",
    pages = "70--81",
    year = "1969"
}

@article{Ong:2020tvo,
    author = "Ong, Yen Chin",
    title = "{Schwinger pair production and the extended uncertainty principle: can heuristic derivations be trusted?}",
    eprint = "2005.12075",
    archivePrefix = "arXiv",
    primaryClass = "gr-qc",
    doi = "10.1140/epjc/s10052-020-8363-2",
    journal = "Eur. Phys. J. C",
    volume = "80",
    number = "8",
    pages = "777",
    year = "2020"
}

@book{Chandrasekhar:1985kt,
    author = "Chandrasekhar, Subrahmanyan",
    title = "{The mathematical theory of black holes}",
    isbn = "978-0-19-850370-5",
    year = "1985"
}

@article{Harko:2009xf,
    author = "Harko, Tiberiu and Kovacs, Zoltan and Lobo, Francisco S. N.",
    title = "{Thin accretion disks in stationary axisymmetric wormhole spacetimes}",
    eprint = "0901.3926",
    archivePrefix = "arXiv",
    primaryClass = "gr-qc",
    doi = "10.1103/PhysRevD.79.064001",
    journal = "Phys. Rev. D",
    volume = "79",
    pages = "064001",
    year = "2009"
}

@article{Errehymy:2025llh,
    author = "Errehymy, A. and Turimov, B. and Syzdykova, A. and Myrzakulov, K. and Alessa, N. and Abdel-Aty, A. -H.",
    title = "{Dymnikova-Schwinger GUP-corrected wormholes in f(R,Lm,T) gravity}",
    doi = "10.1016/j.nuclphysb.2025.117116",
    journal = "Nucl. Phys. B",
    volume = "1019",
    pages = "117116",
    year = "2025"
}

@article{Errehymy:2025djk, 
    author = "Errehymy, A. and Khedif, Y. and Daoud, M. and Myrzakulov, K. and Turimov, B. and Myrzakul, T.",
    title = "{Quantum corrections to Dymnikova-Schwinger black holes in Einstein-Gauss-Bonnet gravity}",
    eprint = "2509.17630",
    archivePrefix = "arXiv",
    primaryClass = "gr-qc",
    doi = "10.1016/j.physletb.2025.139915",
    journal = "Phys. Lett. B",
    volume = "870",
    pages = "139915",
    year = "2025"
}

@article{Estrada:2023pny,
    author = "Estrada, Milko and Muniz, Celio R.",
    title = "{Dymnikova-Schwinger traversable wormholes}",
    eprint = "2301.05037",
    archivePrefix = "arXiv",
    primaryClass = "gr-qc",
    doi = "10.1088/1475-7516/2023/03/055",
    journal = "JCAP",
    volume = "03",
    pages = "055",
    year = "2023"
}

@article{Alencar:2023wyf,
    author = "Alencar, G. and Estrada, Milko and Muniz, C. R. and Olmo, Gonzalo J.",
    title = "{Dymnikova GUP-corrected black holes}",
    eprint = "2309.03920",
    archivePrefix = "arXiv",
    primaryClass = "gr-qc",
    doi = "10.1088/1475-7516/2023/11/100",
    journal = "JCAP",
    volume = "11",
    pages = "100",
    year = "2023"
}

@article{Dymnikova:1992ux,
    author = "Dymnikova, I.",
    title = "{Vacuum nonsingular black hole}",
    doi = "10.1007/BF00760226",
    journal = "Gen. Rel. Grav.",
    volume = "24",
    pages = "235--242",
    year = "1992"
}

@article{Gliner:1966cgu,
    author = "Gliner, E. B.",
    title = "{Algebraic Properties of the Energy-momentum Tensor and Vacuum-like States of Matter}",
    journal = "Sov. Phys. JETP",
    volume = "22",
    pages = "378--382",
    year = "1966"
}

@article{Teo:1998dp,
    author = "Teo, Edward",
    title = "{Rotating traversable wormholes}",
    eprint = "gr-qc/9803098",
    archivePrefix = "arXiv",
    reportNumber = "DAMTP-R-98-17",
    doi = "10.1103/PhysRevD.58.024014",
    journal = "Phys. Rev. D",
    volume = "58",
    pages = "024014",
    year = "1998"
}

\end{document}

	\appendix
	\section{The dragging of the inertial frame\label{appenA}}
	In this appendix, we follow Chandrasekhar's approach to illustrate the presence of a frame-dragging effect in the spacetime described by Eq.~(\ref{s2be1}). Accordingly, the contravariant components of the metric tensor \(g^{\mu\nu}\) can be written in matrix form\footnote{Here, we adopt the notation \(t \rightarrow 0\), \(\theta \rightarrow 2\), and \(r \rightarrow 3\).}.
	\begin{equation}
	    \label{A1}
	    (g^{\mu\nu})=\left(
        \begin{array}{cccc}
        -\frac{1}{N^2} & -\frac{\omega }{N^2} & 0 & 0 \\
        -\frac{\omega }{N^2} & -\frac{A}{K^2 N^2 r^2} & 0 & 0 \\
        0 & 0 & \frac{1}{K^2 r^2} & 0 \\
        0 & 0 & 0 & e^{-\mu } \\
        \end{array}
        \right),
	\end{equation}
	where is defined as 
	\begin{equation}
	    \label{A2}
	    A=(K r \omega -N \csc (\theta )) (K r \omega +N \csc (\theta )).
	\end{equation}
    This spacetime is associated with the following tetrad\footnote{We adopt the same notation as in Ref.~\cite{Chandrasekhar:1985kt}.}.
    \begin{equation}
        \label{A3}
        \begin{aligned}
            e_{(c)\mu}&=(-N,0,0,0),\\
            e_{(1)\mu}&=(-rK\omega\sin\theta, rK\sin\theta,0,0),\\
            e_{(2)\mu}&=(0,0,rK,0),\\
            e_{(3)\mu}&=(0,0,0,e^\frac{\mu}{2}).
        \end{aligned}
    \end{equation}
   Using the relation \(e^{\;\;\;\;\mu}_{(a)}=g^{\mu\nu}e_{(a)\nu}\), the corresponding contravariant vectors can be expressed as follows:
        \begin{eqnarray}
            e^{\;\;\;\;\mu}_{(c)}&=\left(\frac{1}{N},\frac{\omega}{N},0,0\right), \label{Aa4}\\
            e^{\;\;\;\;\mu}_{(1)}&=\left(0, \frac{1}{rK\sin\theta},0,0\right),\label{Ab4}\\
            e^{\;\;\;\;\mu}_{(2)}&=\left(0,0,\frac{1}{rK},0\right),\label{Ac4}\\
           e^{\;\;\;\;\mu}_{(3)}&=\left(0,0,0,e^{-\frac{\mu}{2}}\right).\label{Ad4}
        \end{eqnarray}
    Therefore, for the tetrad so defined, we have 
    \begin{equation}
        \label{A5}
        e^{\;\;\;\;\mu}_{(a)}e_{(b)\mu}=\eta_{(a)(b)}=\left(
        \begin{array}{cccc}
        -1 & 0 & 0 & 0\\
         0 & 1 & 0 & 0\\
         0 & 0 & 1 & 0\\
         0 & 0 & 0 & 1
        \end{array}
        \right).
    \end{equation}
    This indicates that the selected frame is Minkowskian, meaning it locally represents an \textit{inertial frame}. One can easily verify that the components of the line element (\ref{s2be1}) follow directly from the relation:
    \begin{equation}
        \label{A6}
        g_{\mu\nu}=\eta^{(a)(b)}e_{(a)\mu}e_{(b)\nu}.
    \end{equation}
    
    The components of the four-velocity take the following form
    \begin{equation}
        \label{A7}
        \begin{array}{ccc}
            u^0=\frac{dt}{d\lambda}, & u^1=\Omega u^0, & u^\alpha=v^{\alpha} u^0\ . 
        \end{array}
    \end{equation}
    Here, \(\alpha = 2, 3\), \(v^\alpha = dx^\alpha/dt\), and \(\Omega \equiv d\varphi/dt\). In the inertial frame, the components of the four-velocity are determined from the relation:
    \begin{equation}
        \label{A8}
        u^{(a)}=\eta^{(a)(b)}e_{(b)\mu}u^\mu,
    \end{equation}
    from which 
    \begin{eqnarray}
            u^{(c)}&= N u^0,\label{Aa9}\\
            u^{(1)}&= (\Omega -\omega)rK\sin\theta u^0,\label{Ab9}\\
            u^{(2)}&= rKv^2u^0,\label{Ac9}\\
            u^{(3)}&= e^\frac{\mu}{2}v^3u^0.\label{Ad9}
    \end{eqnarray}
    From the second relation in Eqs.~(\ref{Aa9})-(\ref{Ad9}), it follows that a point undergoing circular motion with angular velocity \(\Omega\) in the coordinates \((t, \varphi, \theta, r)\) will appear to rotate with angular velocity \((\Omega - \omega) r K \sin\theta, u^0\) in the inertial frame. Conversely, a point at rest in the local inertial frame (i.e., \(u^{(1)} = u^{(2)} = u^{(3)} = 0)\) will have angular velocity \(\omega\) when viewed in the coordinate frame. This nonzero \(\omega\) therefore represents the dragging of inertial frames. In an asymptotically flat spacetime, this reduces to \(\omega = 2J/r^3\).



\author{B. Turimov\orcidI{}}%
\email[]{bturimov@astrin.uz}
\affiliation{Central Asian University, Milliy bog Str. 264, Tashkent, 111221, Uzbekistan}
\newcommand{\orcidauthorI}{0000-0003-1502-2053} 